\documentclass[12pt,a4paper]{article}
\usepackage{graphicx}
\usepackage{bm}
\usepackage{amssymb}
\begin{document}
\begin{titlepage}



\begin{flushright}
{\bf Budker INP 2002-54\\
September 25, 2002 }
\end{flushright}
\vspace{2cm}






\centerline{\Large {\bf Emission of polarized photons from unpolarized
 }}
 \vskip .25cm
 \centerline{\Large {\bf  electrons moving in crystals  
 \footnote{This work was supported in part
 by the RFBR Grant 01-02-16926} }}
 \vskip .5cm
 \centerline{\large{\bf V.M.Strakhovenko }}
 \vskip .5cm
 \centerline{G.I.Budker Institute of Nuclear Physics,
 630090, Novosibirsk, Russia}
 \vskip 2.0cm
\begin{abstract}
Radiation emitted by unpolarized high-energy electrons penetrating crystals may be linearly polarized. This occurs when the particle velocity makes an angle, with respect to some major crystal axis, being sufficiently larger than the axial-channelling angle. For such orientation, a complete description of spectral and polarization characteristics of the radiation is derived. At planar channelling, a non-perturbative contribution to the probability of the process appears caused by the plane field, and we must solve exactly a one~-~dimensional mechanical problem. For that, the approximate form of the actual  plane potential is suggested which provides a precise fit for any crystal plane and an analytical solution to the motion problem. In a practical case, we must consider electron-photon showers developing in sufficiently thick crystals. For the first time, this development  is described taking into account the polarization of photons. We discuss qualitative features of the phenomenon, present results of numerical calculations for thin and thick crystals, and evaluate the possibility of the use of differently oriented crystals in a polarized
hard photon source.
\end{abstract}
\vspace{2cm}
\end{titlepage}
\newpage
\section{Introduction}
\label{}
High-energy electrons emit the specific radiation while penetrating through single crystals. The shape of the spectrum and the intensity of this radiation depend on the electron energy and crystal orientation. This phenomenon has been widely investigated theoretically and experimentally ( see e.g., \cite{book} and references cited there). Photon beams obtained in such a way may  be used, for example, in photo-production experiments of different kind. In particular, they may serve as $\gamma$ - beams in $\gamma\,\gamma$-colliders. 

In this paper, we focus on the polarization properties of emitted photons as the polarization offers additional opportunities in the experimental study. Initial electrons are assumed to be unpolarized. Then, using crystals, only linear polarization may be obtained. We start with the well known formula derived by means of the quasi-classical operator method ( see e.g., \cite{book}). Remember, that within this method the probabilities of QED processes may be expressed by way of classical trajectories of charged particles, provided that the external field involved satisfies some conditions specified in \cite{quasKS}. Fortunately, the electric field of a crystal responsible for the coherent processes satisfies these conditions. In this stage, we have to solve a two~-~dimensional mechanical problem.
 
Further consideration exploits the approach developed in \cite{method}. As shown in \cite{method}, the mechanical problem mentioned becomes essentially one~-~dimensional for the angles of incidence( w.r.t. some major crystal axis), $\vartheta_0$, which satisfy the condition $ \theta_{sp}\le\vartheta_0 \ll 1$. The magnitude of the angle  $\theta_{sp}$
depends on the electron energy and the axis chosen. It was accurately determined in \cite{method} being sufficiently larger than the characteristic angle for axial channelling . Note that the region $\vartheta_0 < \theta_{sp} $ is not interesting here, as we do not expect any polarization of the radiation from unpolarized electrons at axial channelling. In the angular region of interest, a finite set of the most strong planes containing the axis is important. Moreover, the azimuthal angle domains, where the influence of the planes on a motion should be taken into account exactly, do not overlap. Then we have for the transverse (w.r.t. the axis ) velocity of an electron: ${\bm v} \simeq {\bm v}_{slow}+{\bm v}_{fast}$. In this expansion, the first term represents the exact solution for the transverse velocity in the field of the only one plane of the set. The second term gives the contribution of other planes which can be taken into account using a perturbation theory and, additionally, the rectilinear trajectory approximation. Note that  periods of the one~-~dimensional motion ("slow") are noticeably larger then those in the perturbation term ("fast"). The slow component is present when the velocity is almost aligned with one of the planes. It disappears when the velocity of a particle forms sufficiently large angles with each plane of the set. In this case, the perturbation theory in the whole crystal potential is applicable being the essence of the so called coherent bremsstrahlung (CB) theory. 
It is clear that the transverse motion of a particle is two~-~dimensional but the component which needs the exact calculation is one~-~dimensional. To find the latter, we propose the new approximate form of the actual  plane potential which provides the precise fit for any crystal plane and allows one to find an analytic expression for the trajectory. That is done in Appendix A where also the explicit form of the velocity Fourier-transform is obtained. 

In Section 2, general expressions are derived giving the instantaneous (probabilities per unite time or length) characteristics of the radiation. Then they are analyzed and simplified. We discuss the qualitative features of the phenomenon and present results of numerical calculations for thin crystals. It turns out, that, for the two types of  the azimuthal orientation mentioned above, not only spectra but also polarization distributions are utterly different. In a practical use, sufficiently thick crystals are needed to get a noticeable yield. In this case, we can not neglect the multiple photon emission, their absorbtion due to the $e^+e^- $-pair production, and the radiative energy loss of charged particles. In other words,for thick crystals, we must consider the $e^+e^- \gamma$-shower. Such a consideration is performed in Section 3, using the formulas obtained in Section 2 along with those describing the $e^+e^- $-pair production by polarized photons. So, we describe the shower development, taking into account the polarization of photons for both basic QED-processes involved. Some results of Monte-Carlo simulations are presented for settings used in NA43 ( see \cite{kirsebom}) and NA59 ( see, e.g., \cite{NA59}) experiments at CERN. The consideration performed allows one to decide for the optimal crystal type, thickness, and orientation providing necessary characteristics of a polarized photon source. 
\section{Radiation at fixed particle energy}
Let us start with the well known formula ( see, e.g., Eq.(16.7) in \cite{book} ) which describes spectral, angular, and polarization distributions of photons emitted by  unpolarized electrons (positrons) at given motion:
\begin{equation}\label{photen}
dw_{\gamma}=\frac{\alpha}{(4\pi)^2}\frac{d^3k}{\omega }\frac {\varepsilon} {\varepsilon'}\int dt_1 dt_2
L(t_1,t_2)\exp[i\frac {\varepsilon} {\varepsilon'}(k,x_1-x_2)]\,\,,
\end{equation}
where $\alpha=1/137$ , $ k^{\mu} \equiv (\omega ,\bm{k})$ is the photon momentum, $x^{\mu}_{1,2}\equiv x^{\mu}(t_{1,2}) $, $x^{\mu}(t)=(t,\bm{x}(t)$ , $\varepsilon'=\varepsilon - \omega$, $\varepsilon$ is the electron energy, and
\begin{equation}\label{Ltt}
L(t_1,t_2)=(\bm{e}^*\bm{v}_1)(\bm{e}\bm{v}_2)(\varphi( \varepsilon) +2 )+[(\bm{e}^*\bm{e})(\bm{v}_1\bm{v}_2-1+\gamma^{-2})-(\bm{e}^*\bm{v}_2)(\bm{e}\bm{v}_1)](\varphi( \varepsilon) -2 ).
\end{equation}
Here $\bm{v}_{1,2}\equiv \bm{v}(t_{1,2})$ is the electron velocity on the classical trajectory $\bm{x}(t)$ , $\gamma=\varepsilon/m, m $ is the electron mass,  $\varphi( \varepsilon)=\varepsilon/\varepsilon'+ \varepsilon' /\varepsilon$. If we set $\varphi( \varepsilon)=2 $ ($\omega=0$)in the expression for $L(t_1,t_2)$, the formula (\ref{photen}) will describe the radiation from the scalar (zero-spin) particle ( see e.g., part 9 in \cite{voldbo} or \cite{book} ). This substitution rule holds for all subsequent formulas in this Section.

Remember that the vectors $\bm{e}$ in Eq.(\ref{Ltt}) correspond to the polarization which would be measured by some detector  and, thereby, have nothing to do with the polarization of emitted photons. The latter is described by the matrix, $dw_{ki}$ in the contraction $dw_\gamma=e_i e_k^* dw_{ki}$. We introduce the vector, $\bm{\xi}$ ($|\bm{\xi}|=1 $),which describes the analyzing ability of a detector, by $e_i e_k^*=(1+\bm{\xi}\bm{\sigma})_{ik} /2\,$, where $\bm{\sigma}$ are the Pauli matrices. The matrix $dw_{ki}$ can be presented in the same form: $dw_{ki}=(A+\bm{B}\bm{\sigma})_{ki}/2\,$. Then we have
\begin{eqnarray}\label{Stokes}
dw_{\gamma} &\!\!=&\!\! \frac{1}{2}(A+\bm{B}\bm{\xi})\equiv \frac{A}{2}(1+\bm{\eta}\bm{\xi})
\, ,\quad \bm{\eta}=\frac{\bm{B}}{A}\,,\quad A=dw_{11}+dw_{22}\,\,,\nonumber\\\\
B_1&\!\!=&\!\!dw_{12}+dw_{21}\,,\quad B_2=i(dw_{12}-dw_{21})\,,
\quad B_3=dw_{11}-dw_{22}\,.\nonumber
\end{eqnarray}
In this equation, the Stokes vector of the radiation, $\bm{\eta} $, is defined. The quantity $A$ gives the value of $dw_{\gamma}$ summed up over polarizations.
 
We assume further that the angular divergence of the electron beam is small enough. Then the same is true for arising photon beam provided that electron energy is sufficiently large ($\varepsilon \gg m $), as the emission angle of a photon (w.r.t. the particle velocity ) is typically  of the order of $\gamma^{-1} \ll 1$. In this case, we can choose some axis (z-axis) such that the momenta of all charged particles and photons make small angles,$\vartheta_{ax}$, w.r.t. this axis:$\,\vartheta_{ax} \ll 1$. Let $\bm{a}_{\perp}$ denotes the component of an arbitrary  vector  $\bm{a}$ transverse to the axis chosen. Using relations
$$ v_z \simeq 1-\frac{1}{2} ( \gamma^{-2} + \bm{v}^2_{\perp})\, ,\, e_z=-(\bm{k}_{\perp}\bm{e}_{\perp})/k_z \simeq (\bm{n}_{\perp}\bm{e}_{\perp})\,,\, n_z \simeq 1-\frac{1}{2}\bm{n}^2_{\perp}\, ,\, \bm{n}=\bm{k}/\omega \,,$$ we can remove z-components of all the vectors from Eq. (\ref{photen}). As a result, we have
\begin{eqnarray}\label{phottr}
dw_{\gamma} &\!\!=&\!\! \frac{\alpha \omega d \omega d^2 n_{\perp}}{(4\pi)^2}\frac {\varepsilon} {\varepsilon'}\int dt_1 dt_2 L(t_1,t_2)\exp(-iD)\,\,,\nonumber\\ \nonumber\\
 D&\!\!=&\!\! \frac {\omega \varepsilon}{2 \varepsilon'}\int\limits ^{t_2}_{t_1} dt [\,\gamma^{-2}+(\bm{n}_{\perp}-\bm{v}_{\perp}(t))^2\,]\,,\nonumber\\\\
L(t_1,t_2)&\!\!=&\!\!(\bm{e}^*_{\perp},\bm{v}_{1\perp}-\bm{n}_{\perp})(\bm{e}_{\perp},
\bm{v}_{2\perp}-\bm{n}_{\perp})(\varphi( \varepsilon) +2 )\nonumber\\\nonumber\\&\!\!-&\!\![(\bm{e}^*_{\perp},\bm{v}_{2\perp}-\bm{n}_{\perp})(\bm{e}_{\perp},
\bm{v}_{1\perp}-\bm{n}_{\perp})+\frac {1}{2}(\bm{e}^*_{\perp} \bm{e}_{\perp})(\bm{v}_{2\perp}-\bm{v}_{1\perp})^2](\varphi( \varepsilon) -2 )\,.\nonumber
\end{eqnarray}
No restrictive conditions (like $(\bm{n} \bm{e})=0 $) are imposed on the vectors $\bm{e}_{\perp}$. This allows one, first, to utilize any fixed basis for the description of the photon polarization regardless of  $\bm{n}$. In what follows, we use the Cartesian basis ($\bm{e}_x,\bm{e}_y,\bm{e}_z $), where $\bm{e}_z $ is directed along the nearest  (making a small angle w.r.t. the particle velocity ) major crystal axis  , $\bm{e}_x $ is within some crystal plane containing this axis, and  $\bm{e}_y $ is perpendicular to this plane. Secondly, the vectors $\bm{e}_{\perp}$ can be considered constant when integrating in (\ref{phottr}) over photon emission angle ( over $ d^2 n_{\perp}$ ). Performing this integration, we obtain the matrix, $w_{ij}^{sp}$, describing the spectral distribution of a radiation:
\begin{eqnarray}\label{spectr}
dw_{ij}^{sp}&\!\!=&\!\! \frac {i\alpha d\omega}{4\pi\gamma^2}\int\limits_{-\infty} ^{\infty}dt\int\limits_{-\infty} ^{\infty}\frac{d\tau}{\tau-i0}\,\exp [-i\lambda\tau(\,1+\rho(t,\tau))]\,L_{ij}\,,\nonumber\\\\
L_{ij}&\!\!=&\!\!\delta_{ij}^{\perp}\Bigl[\frac{1}{4}(\varphi( \varepsilon) -2 )(\bm{g}_2-\bm{g}_1)^2+\frac{i}{\lambda(\tau-i0)}\Bigr]+\frac{\varphi( \varepsilon)}{2} \Bigl(g_{2i}g_{1j}-g_{2j}g_{1i}\Bigr)-\Bigl(g_{2i}g_{1j}+g_{2j}g_{1i}\Bigr)\,,\nonumber
\end{eqnarray}
where $t=(t_1+t_2)/2\, $,\quad $\tau=t_2-t_1\,$,\quad $\lambda=\omega m^2/(2\varepsilon \varepsilon')\,$, \quad $\bm{g}_{1,2}\equiv \bm{g}(t_{1,2})\, $,\quad and $$\bm{g}(s)=\gamma\Bigl[\,\bm{v}_{\perp}(s)- \frac{1}{\tau}\int\limits_{-\tau/2} ^{\tau/2}dx\,\bm{v}_{\perp}(t+x) \Bigr]\,,\qquad \rho(t,\tau)=  \frac{1}{\tau}\int\limits_{-\tau/2} ^{\tau/2}dx\,\bm{g}^2(t+x). $$

In further consideration, we use the approach developed in \cite{method}. Recollect that
the transverse velocity can be represented as the sum:  $\bm {v}_{\perp} \simeq \bm {v}^{slow}_{\perp}+\bm {v}^{fast}_{\perp}$, where $\bm {v}^{fast}_{\perp}$  given by Eq.(3) in \cite{method} is characterized by  small amplitudes and large frequencies. On the contrary, large amplitudes and small frequencies are inherent in the term $\bm {v}^{slow}_{\perp}$ which corresponds to the one~-~dimensional motion in the field of a plane. Its explicit form reads  $\bm {v}^{slow}_{\perp}(t)=\bm{e}_y \displaystyle {\sum_n}\,v_n \exp(in\omega_0 t)\,$. Here $\omega_0\,$ is the frequency of this motion, and $v_n\,$ is the velocity Fourier transform calculated in Appendix A. Correspondingly, the quantity $\bm{g}(s)\,$ in (\ref{spectr}) turns into the sum: $\bm{g}(s)=\bm{l}(s)+\bm{w}(s)\,$ where
\begin{eqnarray}\label{lws}
\bm{l}(s)\!\!&=&\!\bm{e}_y\,\sum\limits_{n}{}(\gamma v_n)\Bigl[e^{in\omega_0 s}- \frac{\sin(n\omega_0 \tau/2)}{n\omega_0 \tau/2}e^{in\omega_0 t}\Bigr]\,,\nonumber\\\\
\bm{w}(s)\!\!&=&\!- \widetilde{\sum\limits_{\bm{q}_{\perp}}^{}}\frac{G(\bm{q}_{\perp}) \bm{q}_{\perp}}{mq_{\parallel}}\Bigl[e^{iq_{\parallel} s}-\frac{\sin(q_{\parallel} \tau/2)}{q_{\parallel} \tau/2}e^{iq_{\parallel} t}\Bigr]\,e^{i(\bm{q}_{\perp} \bm{\rho}_0)}   \,.\nonumber
\end{eqnarray}
Here $\bm{q}$ are discrete reciprocal lattice vectors, $ G(\bm{q})\,$ are coefficients in the fourier series presenting the crystal potential ( see Chapter 9 of \cite{book} for the explicit form of $\bm{q}$ and $G(\bm{q})\,$ ), and $q_{\parallel}=(\bm{q}_{\perp}\bm{v}) \,$. The tilde in the expression for $\bm{w}(s)$ means that the sum does not contain $\bm{q}_{\perp}\,\parallel \,\bm{e}_y\,$ which just form the plane potential. 

In principle, using (\ref{lws}), one can perform the integration over $t\,$ and $\tau $ straight in (\ref{spectr}). However, as we have already obtained the quantity  $\bm{w}(s)$ by means of some perturbation procedure, it is more consequently to continue in the same way. So, as in \cite{method}, we expand the exponential function in (\ref{spectr}) in $\bm{w}(s)\,$, keeping quadratic terms, and obtain 
\begin{equation}\label{expsp}
dw_{ij}^{sp}= \frac {i\alpha d\omega}{4\pi\gamma^2}\int\limits_{-\infty} ^{\infty}dt\int\limits_{-\infty} ^{\infty}\frac{d\tau}{\tau-i0}\,\exp [-i\lambda\tau(\,1+\rho_{slow}(t,\tau))]\,\sum\limits_{n=1}^{3}\,C_{ij}^{(n)}\,,
\end{equation} 
where $\rho_{slow}(t,\tau)= \tau^{-1}\int_{-\tau/2}^{\tau/2}dx\,\bm{l}^2(t+x)\,$ and matrices $C_{ij}^{(n)}$ are
\begin{eqnarray}\label{matrcij}
C_{ij}^{(1)}\!\!&=&\!\delta_{ij}^{\perp}\Bigl[\frac{i}{\lambda(\tau-i0)}+\frac{1}
{4} (\varphi( \varepsilon) -2 )(\bm{l}_2-\bm{l}_1)^2\Bigr]-l_{1i}l_{2j}-l_{1j}l_{2i}\,,\nonumber\\ \nonumber\\
C_{ij}^{(2)}\!\!&=&\!\widetilde{\sum\limits_{\bm{q}_{\perp}}^{}}\Biggl | \frac{G(\bm{q}_{\perp})\bm{q}_{\perp}}{mq_{\parallel}}\Biggr |^2\Biggl \{ \delta_{ij}^{\perp}\Bigl[1-f^2(\zeta)+(\varphi( \varepsilon) -2 )\sin ^2\zeta\Bigr]-\nu_i\nu_j\Bigl[f_{(+)}^2(\zeta)+f_{(-)}^2(\zeta)\Bigr]\Biggr\}\,,\nonumber\\ \nonumber\\
C_{ij}^{(3)}\!\!&=&\!2i\lambda \widetilde{\sum\limits_{\bm{q}_{\perp}}^{}}\Biggl | \frac{G(\bm{q}_{\perp})\bm{q}_{\perp}}{mq_{\parallel}}\Biggr |^2\,J_{ij}\,,\\
J_{ij}\!\!&=&\!\frac{\tau}{2}\Bigl (1-f^2(\zeta)\Bigr )\Bigl[\delta_{ij}^{\perp}\frac{i} {\lambda(\tau-i0)}- C_{ij}^{(1)}\Bigr]+i\lambda \bigl | \Phi \bigr |^2C_{ij}^{(1)}+ \Phi\Biggl \{    i \delta_{ij}^{\perp}(\varphi( \varepsilon) -2 )(\bm{l}_2-\bm{l}_1,\bm{\nu})\sin \zeta\nonumber\\ \nonumber\\
\!\!&+&\!(l_{1i}\nu_{j}+l_{1j}\nu_{i})f_{(-)}(\zeta)+(l_{2i}\nu_{j}+l_{2j}\nu_{i})f_{(+)}
(\zeta)+\frac{\varphi( \varepsilon)}{2}\Bigl[(l_{1i}\nu_{j}-l_{1j}\nu_{i})f_{(-)}(\zeta) \nonumber\\ \nonumber\\
\!\!&-&\!(l_{2i}\nu_{j}-l_{2j}\nu_{i})f_{(+)}(\zeta)\Bigr] \Biggr\}\,,\quad \Phi= \int \limits_{-\tau/2} ^{\tau/2}dx\,(\bm{\nu}\bm{l}(t+x))\Bigl[e^{iq_{\parallel}x}-f(\zeta)\Bigr]
\,,\nonumber\\ f(x)\!\!&=&\!\frac{\sin x}{x}\,,\quad f_{(\pm)}(x)=e^{\pm ix}- f(x)\,,\quad
\bm{\nu}=\frac{\bm{q}_{\perp}}{|\bm{q}_{\perp}|}\,,\quad \zeta=q_{\parallel}\tau/2\, .\nonumber
\end{eqnarray}
Using Eqs. (\ref{expsp}),(\ref{matrcij}), we obtain the expression for the probability of a photon detection irrespective of its polarization, $dw_{\gamma}^{unp}=dw_{ij}^{sp} \delta_{ij}^{\perp}\,$, wich coincides with Eq.(4) in \cite{method}.

Let us dwell on the plane field contribution (PFC) to the radiation. This is given by the term in (\ref{expsp}) which is proportional to $C_{ij}^{(1)}$. To make estimates, we should recollect some properties of the quantity $\rho_{slow}(t,\tau)$ which enters into the phase, $\Psi(\tau) = \lambda\tau(\,1+\rho_{slow}(t,\tau))$. So, $\rho_{slow}(t,\tau)$ is the even,
positive, and monotonically increasing function of $\tau\,$. It satisfies  the inequality $\rho_{slow}(t,\tau)<\rho (\infty)<\rho_c =U_{pl}\varepsilon/m^2 $ where $U_{pl}\,$ is the potential well depth of the plane and $\rho(\infty)$ is the limiting value of  $\rho_{slow}(t,\tau)$ as $ |\tau | \rightarrow \infty$ . Using (\ref{lws}), we can easily estimate the behavior  of $\rho_{slow}(t,\tau)$ at large and small $\tau\,$ 
\begin{equation}\label{roexpa}
\rho(\infty)=\sum \limits_{n \ne 0}^{}|\gamma v_n |^2=\frac{<p_y^2>- <p_y>^2  }{m^2}\,,\quad \rho(\omega_0 \tau \ll 1)\simeq \frac {1}{12}(\gamma \dot {v}_y(t)\tau)^2\,,
\end{equation} 
where $<\ldots>$ means time averaging, $p_y$ and $\dot  {v}_y\,$ are correspondingly y-components of the  momentum and acceleration of a charged particle. Let $\vartheta_0$ and $\phi_0 $ be the polar and azimuthal angles of incidence counted off  from the axis and plane correspondingly. Then the velocity of a particle makes the angle, $\psi$, w.r.t. the plane, at that, $\sin \psi= \sin \vartheta_0 \sin \phi_0 $ or, as $\vartheta_0 \ll 1$ , we have  $ \psi \simeq  \vartheta_0 \sin \phi_0 $. At channelling ( $ \psi < \theta_{pl}= (2 U_{pl}/\varepsilon)^{1/2}\,$) , we have $\rho (\infty)\lesssim \rho_c$. At above-barrier motion ( $ \psi > \theta_{pl}\,$), the quantity $\rho (\infty)$ decreases fast being of the order of $0.1(U_{pl}/m\psi )^2\,$ for $ \psi \gg \theta_{pl}\,$ (see, e.g., p.430 in \cite{book}).

For $\rho (\infty)\ll 1\,$ , we can expand the exponential function in (\ref{expsp}) in 
powers of $\rho_{slow}(t,\tau)$ retaining only the linear term ( dipole approximation). In this case, the slow-fast interference term, $C_{ij}^{(3)}\,$, gives higher-order corrections and should be neglected, while $C_{ij}^{(1)}\,$ after averaging over time $t$ reads
\begin{equation}\label{c1small}
C_{ij}^{(1)}(\rho \ll 1)=\sum\limits_{n}^{}|\gamma v_n |^2\Biggl \{ \delta_{ij}^{\perp}\Bigl[1-f^2(\zeta)+(\varphi( \varepsilon) -2 )\sin ^2\zeta\Bigr]-e_{yi}e_{yj}\Bigl[f_{(+)}^2(\zeta)+f_{(-)}^2(\zeta)\Bigr]\Biggr\}\,,
\end{equation} 
where $\zeta=n\omega_0 \tau/2\,$ , $e_{yi}=\delta_{2i} $. This expression has the same structure as  $C_{ij}^{(2)}$ ( cf. (\ref{matrcij})) since both were obtained by means of the perturbation theory. For $\rho_c \ll 1\,$ , Eq.(\ref{c1small}) is valid at any value of the angle $\psi$. In the opposite case when $\rho_c \gtrsim 1\,$ , it holds for $\psi \gg \theta_{pl}$. The latter condition provides also the applicability of the rectilinear trajectory approximation in the  calculation of $v_n$ and $\omega_0$. So, as we have checked using the explicit form of $v_n$ and $\omega_0$ ( see Appendix A ), for $\psi \gg \theta_{pl}$ , $C_{ij}^{(1)}$ and $C_{ij}^{(2)}$ are essentially the same except that the summation is performed in $C_{ij}^{(1)}$ over the subset of $\bm{q}_{\perp}$ parallel to $\bm{e}_y$ , while in $C_{ij}^{(2)}$ the complementary subset is used. As a result, when the particle velocity is well off all major planes, the spectrum and polarization of a radiation is described by the term $C_{ij}^{(2)}$ if we extend the summation in it over all $\bm{q}_{\perp} $. Then it reproduces results of the so-called coherent bremsstrahlung (CB) theory. In \cite{uspekhi}( see Eqs. (4.8),(1.5), the correlation was explained of this theory with the Compton scattering (off electrons) of the equivalent photons representing in a proper reference frame the periodic crystal field.

For $\rho (\infty)\gg 1\,$ , the constant field approximation (CFA) is widely used. It can be applied when the magnitude of the external ( not uniform ) field is almost constant on the particle trajectory during the radiation formation time, $\tau_f\,$. The latter is determined by the condition $\Psi(\tau_f)\lesssim 1\,$. In our case, the applicability condition of CFA reads $\omega_0 \tau_f \ll 1\,$. As  $\tau_f\,$ depends on photon energy, 
$\omega\,$ , this condition can not be fulfilled everywhere in the radiation spectrum. In particular, it is violated for $\omega \lesssim 2\gamma^2\omega_0/\rho (\infty)$. The power spectrum is maximum  at $u\equiv \omega/(\varepsilon-\omega)\sim \chi = \gamma^2 \dot{v}(t)/m\,$ . Using Eq. (\ref{roexpa}), we find at $u \gtrsim \chi \,$ for the formation time $\tau_f \sim 2\omega_0/(\gamma \dot{v}(t)\,$. Numerical estimates of the latter expression show that, even in this part of the spectrum, the CFA becomes valid at rather high electron energy. For example, it happens at $\varepsilon \gg 100$ GeV  for $(110)$-plane of diamond and silicon, and at $\varepsilon \gg 10$ GeV  for the same plane of tungsten. Using CFA, the probability, $dw_{\gamma}$, was calculated in \cite{hardph}. We emphasize that no approximations are used in the present paper at the PFC calculation. However, the results of \cite{hardph} are useful here since they allow us to check the procedure of the exact calculation and estimate the magnitude of the interference term, $C_{ij}^{(3)}\,$, as well as the influence of the slow motion on the CB term $C_{ij}^{(2)}\,$. This influence is due to the presence of $\rho_{slow}(t,\tau)\,$ in the phase $\Psi(\tau)\,$. According to \cite{hardph}, the scale of the effect is given by the parameter $\mu=\chi/s(q_{\parallel})$ where $s(q_{\parallel})=2\varepsilon  |q_{\parallel}| /m^2\,$ is a conventional kinematic parameter in the Compton scattering . Remember, that the power spectrum of the latter is maximum just at the kinematic boundary, $u=s(q_{\parallel})$. So that, the plane field and CB contributions to the radiation intensity are well separated for $\mu \ll 1\,$.  The parameter $\mu\,$ is essentially the deviation of the particle velocity due to the external field action during the hard-photon formation time, $\tau_h \sim |q_{\parallel}|^{-1}\,$ , measured in units of the characteristic emission angle $\theta_{ph}=\gamma^{-1}$. For given plane and substance, this parameter depends only on the angle of incidence, $\vartheta_0\,$ , namely, $\mu= C_{\mu}/\vartheta_0$(mrad). Using the Tables 9.2 and 15.1 in \cite{book} , we obtain for $<110>$ planes, $C_{\mu}$(diamond)$ \simeq C_{\mu}$(silicon)$ \simeq 0.01\,$, and  $C_{\mu}$(tungsten) $\simeq 0.06\,$. For weaker planes of the same crystal, $C_{\mu}$ is smaller being proportional to the magnitude of the plane electric field. From Fig.1 in \cite{hardph}, the influence of the field on the Compton scattering can be neglected if $\mu $ does not exceed several hundredth. In turn, the interference term provides corrections to PFC of the order of $\rho_{fast}\simeq \widetilde{\sum} | G(\bm{q}_{\perp})\bm{q}_{\perp}/(mq_{\parallel}) |^2\,$ which, by definition , is much smaller than unity. Assuming $\mu < 0.05\,$ , we obtain from (\ref{expsp}) a simple but still rather accurate expression for the photon emission probability, $ dW $, per unite length (time) as well as the spectral distribution of polarization. According to the analysis performed in \cite{hardph}, the accuracy of 5 $\%$ or better is expected. Let us present $ dW $ in the form introduced in (\ref{Stokes}) as just this form will be used in further calculations: 
\begin{equation}\label{prosp}
\frac{dW}{d\omega}=\frac{dW^{(F)}}{d\omega}+ \frac{dW^{(C)}}{d\omega}\,,\quad \frac{dW^{(C,F)}}{d\omega} =\frac{1}{2}\Bigl ( A^{(C,F)}+\bm{B}^{(C,F)}\bm{\xi} \Bigr)\,,
\end{equation}  
where the superscripts $C,F$ are correspondingly for the CB contribution and PFC.    For the latter (plane field contribution), we have
\begin{eqnarray}\label{CPF}
\Bigl ( A^{(F)},\bm{B}^{(F)}\Bigr)\!\!&=&\!\!\frac{\alpha}{\pi^2\gamma^2}\int\limits _{0}^{\infty}\frac{ds}{s}\int\limits _{0}^{\pi}dx \Bigl ( a^{(F)},\bm{b}^{(F)}\Bigr)\,;\, b^{(F)}_1=b^{(F)}_2=0\,, \, b^{(F)}_3=\Bigl ( D^2_2-D^2_1\Bigr)\sin \Psi\,,\nonumber \\ \nonumber\\
a^{(F)}\!\!&=&\!\!\Bigl [ \bigl( \varphi(\varepsilon)-1\bigr)D^2_1-D^2_2\Bigr ]\sin \Psi + \frac{\cos(s\delta)-\cos \Psi}{s\delta}\,,\quad \Psi=s\delta\bigl(1+D_3 \bigr)\,,\nonumber \\\\D_1\!\!&=&\!\! 2\gamma \sum\limits_{n=1}^{\infty}v_n\sin(ns)\sin(nx)\,,\quad D_2=2\gamma \sum\limits_{n=1}^{\infty}v_ng(ns)\cos(nx)\,, \nonumber \\ \nonumber\\D_3\!\!&=&\!\! 2\gamma^2 \sum\limits_{n,m=1}^{\infty}v_n v_m \bigl \{\bigl [f\bigl((n+m)s\bigr)-f(ns)f(ms)\bigr ]\cos\bigl((n+m)x\bigr)+\bigl(m\rightarrow -m\bigr)\bigr \}\,,\nonumber \\ \nonumber\\\delta\!\!&=&\!\! \frac{m^2u} {\varepsilon \omega_0}\,,\quad f(x)=\frac{\sin x}{x}\,,\quad g(x)=\cos x-f(x)\,,\nonumber
\end{eqnarray}
where the integration over $x=\omega_0 t$ corresponds to time averaging, the variable $s=\omega_0 \tau/2$ is used instead of $\tau\,$. When the influence of the plane field on the Compton scattering is neglected ( the factor $\exp (-i\lambda\tau \rho_{slow}(t,\tau)) $ is omitted in the term $\propto C_{ij}^{(2)}$ in (\ref{expsp})), the integrals over $\tau\,$ are easily taken and we obtain for the CB contribution:
\begin{eqnarray}\label{ccom}
\Bigl ( A^{(C)},\bm{B}^{(C)}\Bigr)\!\!&=&\!\!\frac{\alpha}{\gamma^2}\widetilde{\sum\limits_{\bm{q}_
{\perp}}^{}}\Biggl | \frac{G(\bm{q}_{\perp})\bm{q}_{\perp}}{mq_{\parallel}}\Biggr |^2\Bigl ( a^{(C)},\bm{b}^{(C)} \Bigr)\theta(1-\beta)\,;\quad \beta=\frac{u}{s(q_{\parallel})}\equiv \frac{\omega m^2}{2\varepsilon \varepsilon' |q_{\parallel}|}\,,\nonumber \\\\a^{(C)}\!\!&=& \!\!\frac{1}{4}\varphi(\varepsilon)-\beta (1-\beta)\,,b^{(C)}_1=\beta^2\nu_1\nu_2 \,,b^{(C)}_2=0\,,b^{(C)}_3=\frac{1}{2}\beta^2(\nu_1^2-\nu_2^2)\,,\nonumber 
\end{eqnarray}              
where $\theta(1-\beta)$ is the step function: $\theta(x)=1\,$ for $x>0$ and $\theta(x)=0\,$ for $x<0\,$ , $\bm{\nu}$ is defined in (\ref{matrcij}).

Using formulas obtained, we present now some examples illustrating the characteristics of 
\begin{figure}[h]
\centering
\includegraphics[width=0.48\textwidth]{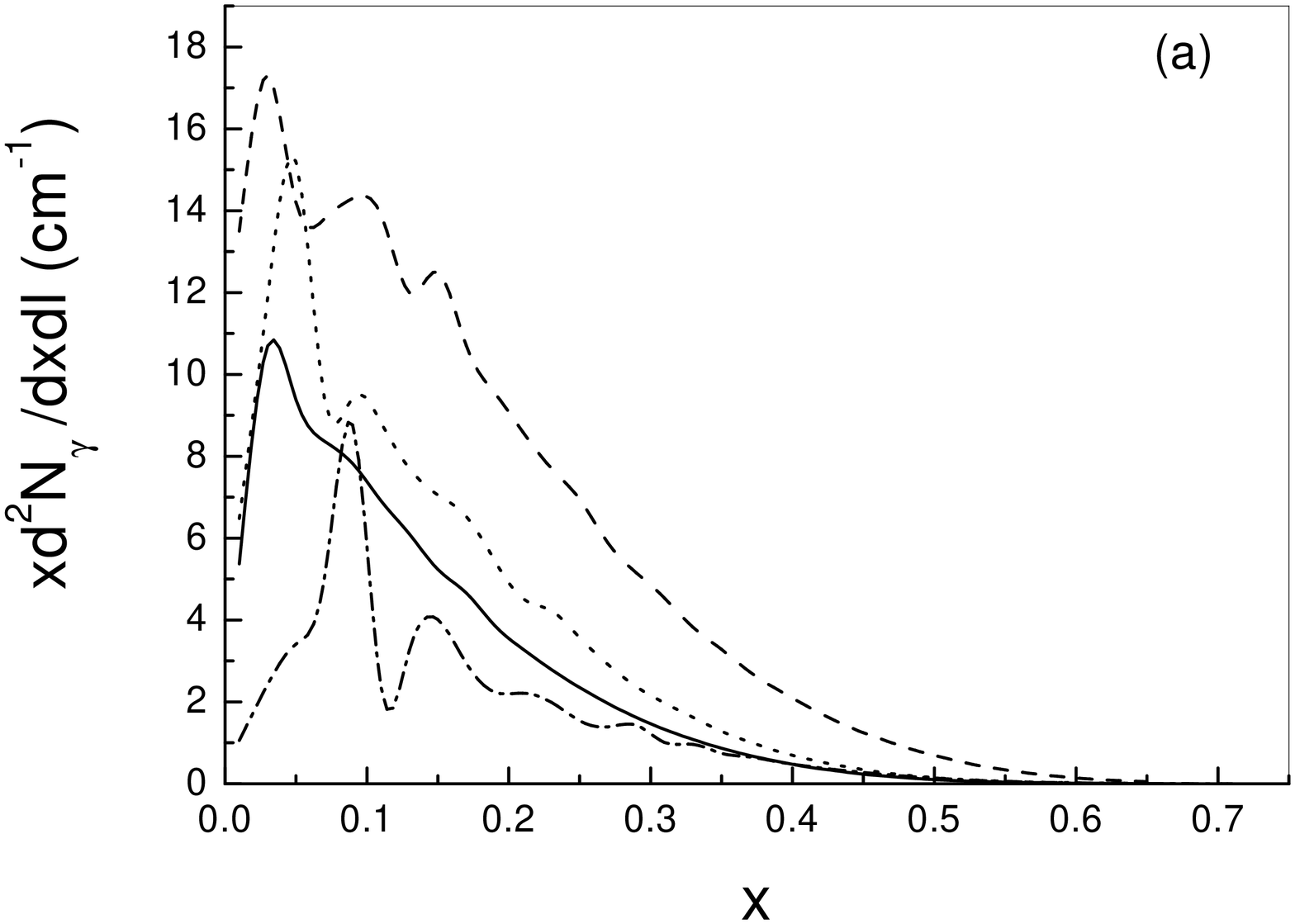}\hspace{0.025\textwidth}
\includegraphics[width=0.48\textwidth]{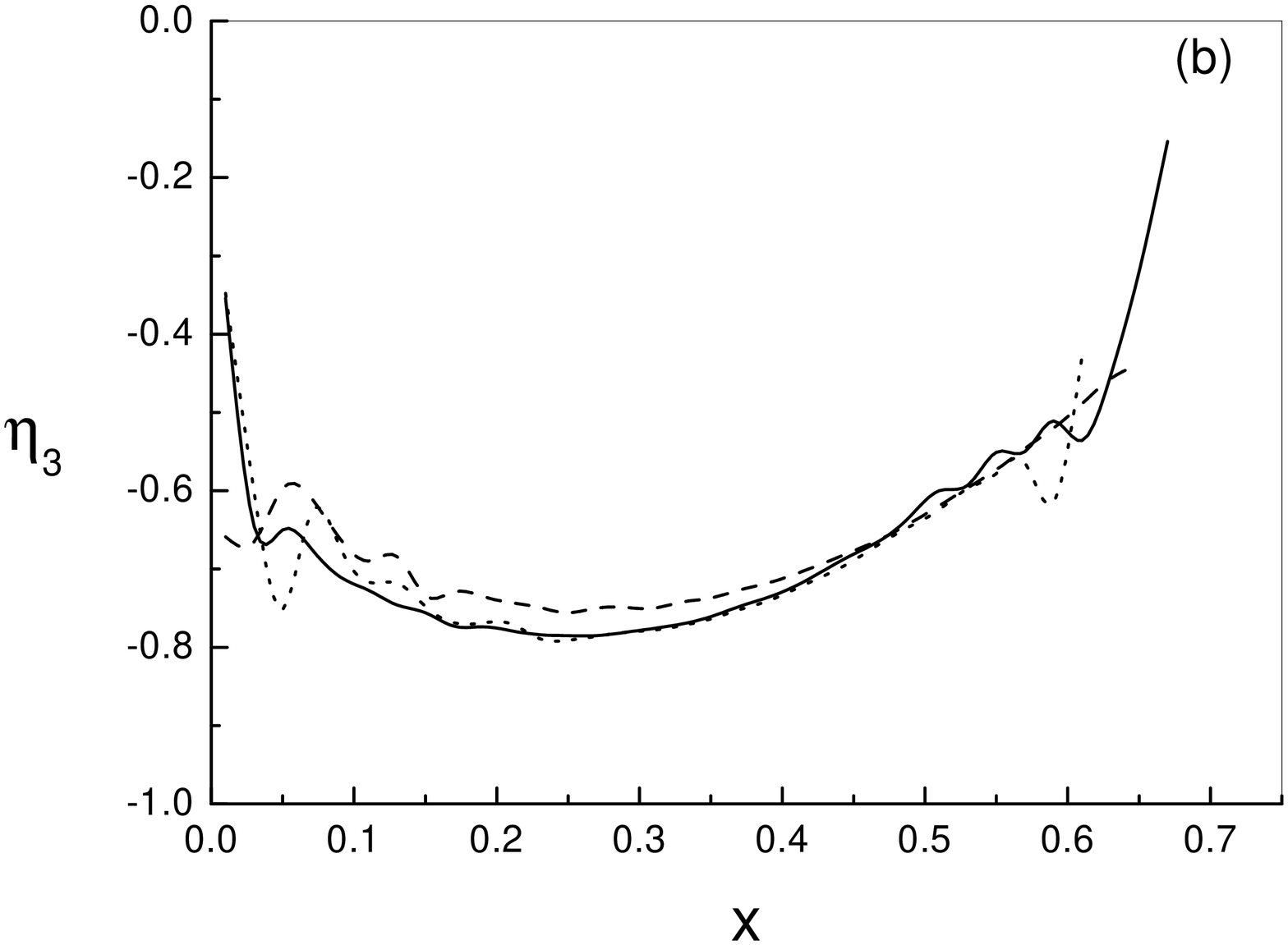}
\caption{(a):intensities $ x d^2N_{\gamma}/dxdl$ at $\varepsilon=180$ GeV for (110)-plane 
   of silicon at $z=-0.25$ (solid), at $z=1$ (dash-dotted); the same in diamond at $\varepsilon=180$ GeV  , $z=-0.25$ for (110)-plane (dashed) and for (001)-plane (dotted). (b): third component of the Stokes vector for these conditions except  $z=1$ ; $z=\varepsilon_{\perp}/U_{pl}$ , $x=\omega/\varepsilon$.}\label{Fig:plane1}
\end{figure}
a radiation. Let us start with the PFC described by (\ref{CPF}). In Fig.\ref{Fig:plane1} , the radiation intensities (a) and polarizations (b) are plotted as functions of $x=\omega/ \varepsilon $. Remember, that the quantities $v_n $ and $\omega_0 $ in (\ref{CPF}) depend on the integral of motion, $\varepsilon_{\perp}$ and so does the radiation. From Fig. \ref {Fig:plane1} , the radiation at channelling ($z=\varepsilon_{\perp}/U_{pl}<0$) is softer and more intensive than that at above-barrier motion ($z>0$). At given $z$ ,the intensity is smaller for a weaker plane (cf. curves for diamond). As expected, the polarization at such one~-~dimensional motion is directed perpendicular to the plane ($\eta_3 <0\,,\,\eta_1= 0$). The polarization  degree is rather high and does not reveal a sharp dependence on $x$. 

\begin{figure}[h]
\centering
\includegraphics[width=0.48\textwidth]{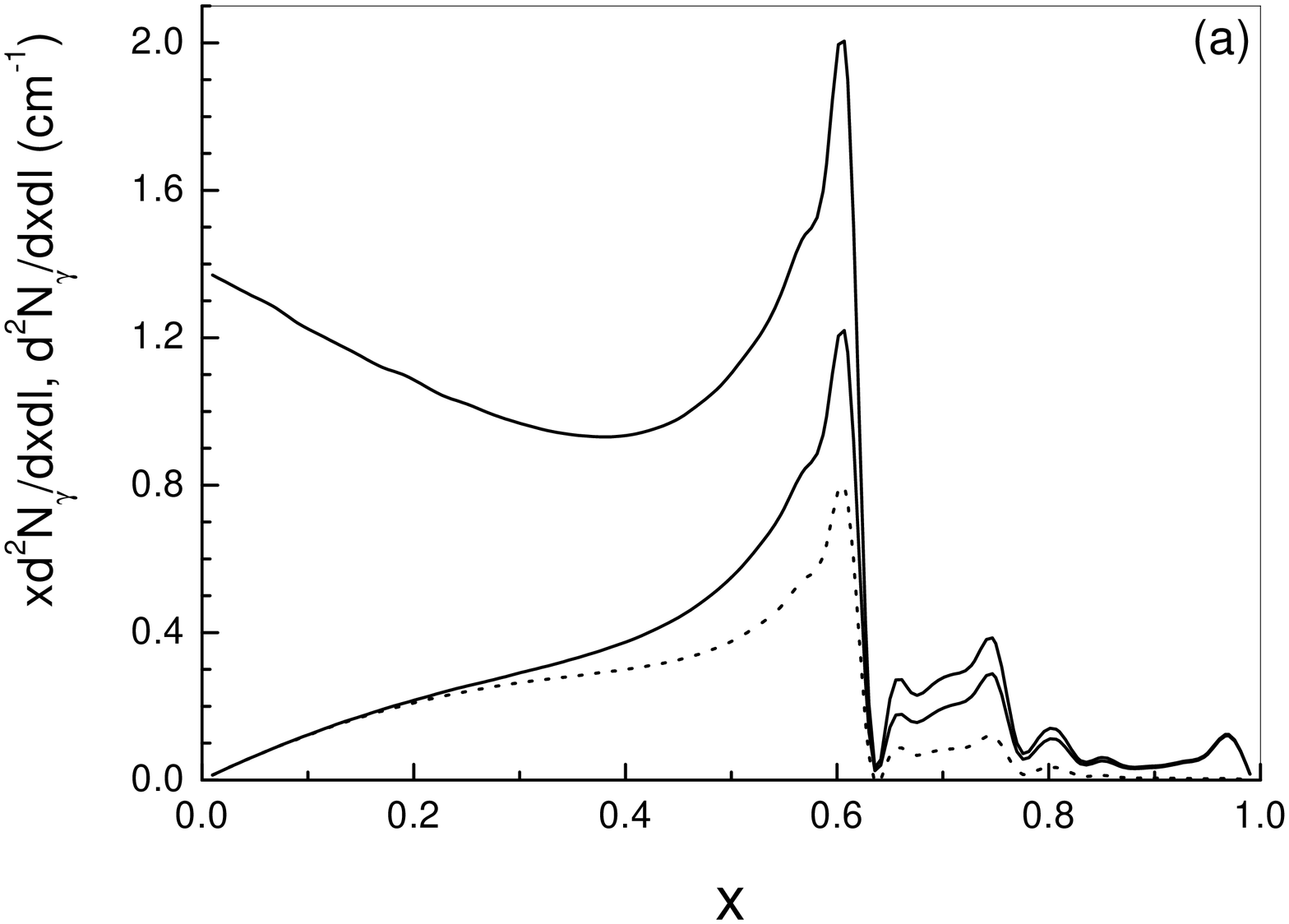}\hspace{0.025\textwidth}
\includegraphics[width=0.48\textwidth]{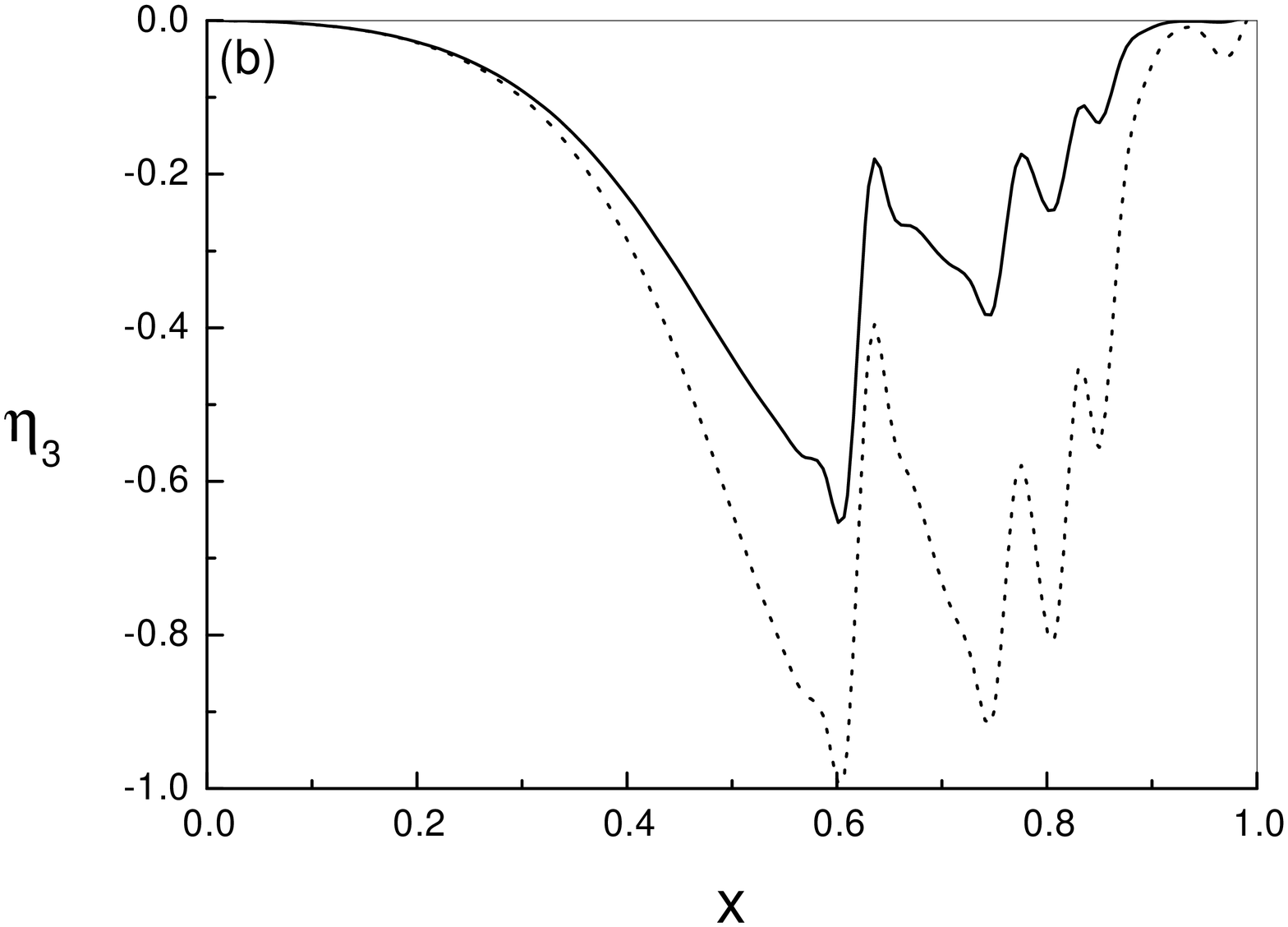}
\caption{ (a): solid curves present intensity $ x d^2N_{\gamma}/dxdl$ and probability $  d^2N_{\gamma}/dxdl$ (upper curve) at $\varepsilon=180$ GeV of CB in silicon ( 5 mrad off the
$<001>$-axis and 180 $\mu$rad off the $(1\bar{1}0)$-plane  ), dotted curve  presents intensity for a "scalar" electron ; (b): third component of the Stokes vector at the same settings}\label {Fig:cbins2}
\end{figure}

In Fig.\ref{Fig:cbins2} , the radiation intensity, probability, and polarization are shown for the angles of incidence $\vartheta_0=5$ mrad and $\phi_0=0.036 $ when $\psi=180 \mu$rad. For the (110)-plane of Si at $\varepsilon=180$ GeV, the channelling angle is of $\theta_{pl}\simeq 15 \mu$rad,i.e., $\psi \gg \theta_{pl}$ ($z\simeq 140 $). Then, as explained above, the first term (PFC)in (\ref{prosp}) has the same form as the second one and we are dealing with pure CB. This term is dominant when, as in our example, $\phi_0\ll 1 $. That is due to the relative smallness of $q_{\parallel}^{(1)}$ from the first subset  (roughly  $q_{\parallel}^{(1)}\sim q_{\parallel}^{(2)}\phi_0$). As a result, the second term has a much smaller amplitude ( $q_{\parallel}^{-2}$ in a partial flux of equivalent photons ) and much higher photon frequencies ($\beta \propto q_{\parallel}^{-1}$ in  (\ref{ccom})). So, it can be neglected for $\phi_0\ll 1 $ and the expression (\ref{ccom}) is reduced to a one~-~dimensional sum. Within this accuracy, we obtain for $\eta_3$ at the maximum of the first harmonic,i.e., at $u=s$ :
\begin{equation}\label{eta3}
\eta_3 (u=s)=-\Bigl \{ \sum\limits_{n=1}^{\infty}\Bigl |\frac{G(nq)}{n}\Bigr |^2 \Bigr\} \diagup\Bigl \{ \sum\limits_{n=1}^{\infty}\Bigl |G(nq)\Bigr |^2 \Bigl[1+\frac{s^2}{2(1+s)} -\frac{2}{n}\Bigl(1-\frac{1}{n} \Bigr)   \Bigr]\Bigr\}\,, 
\end{equation}
where $s=2\varepsilon q\psi/m^2\,\,,\, q=2\pi/d_{pl}\,,$ and $d_{pl}$ is the inter-planar distance. The contribution of the first harmonic ($n=1$) to $\eta_3$ at $u=s$ is $\eta_3^{(1)}=-\bigl [1+s^2/(1+s)/2\bigr]^{-1}$. Worthy to note that $\eta_3^{(1)}$ is independent of $G(q)$. In our case, when $s^2/(1+s)/2\simeq 0.49$ , it overestimates the exact value  (\ref{eta3}) by $8\%$. Recollect now, that according to \cite{hardph} each equivalent photon is completely linearly polarized along its $\bm{q}_{\perp}$. In the first subset, all such photons have $\bm{q}_{\perp}$ perpendicular to the plane. Thus, the whole equivalent photon beam produced by this subset is completely linearly polarized leading to the polarization of emitted radiation perpendicular to the plane ($\eta_3 <0\,,\,\eta_1= 0$). Actually, the second term in $dW/d\omega$ (see (\ref{prosp})) has not been neglected in our calculation giving at this orientation non~-~vanishing but extremely small value of $\eta_1$.
To give a glimpse of a role of the electron spin, the intensity and polarization from a "scalar" (zero spin) electron are presented in Fig.\ref{Fig:cbins2}(dotted curves). The spin terms in (\ref{Ltt}) and subsequent formulas are proportional to $\varphi(\varepsilon)-2= \omega^2/(\varepsilon \varepsilon')$. Therefore a difference in radiation characteristics becomes observable at sufficiently large photon energies ( for $x>0.3$ in Fig.\ref {Fig:cbins2}). The peaks do not move being determined, at given orientation, solely by the particle energy, $\varepsilon$ . Due to the absence of the spin terms, the radiation from a "scalar" electron is less intensive and, because of that, has a higher polarization. Really, the quantity $\bm{B}^{(C)}$ in (\ref{ccom}) is independent of the particle spin, while $A^{(C)}$, which is proportional to the intensity, appears in the denominator of the equation defining a polarization ( $\bm{\eta}=\bm{B}/A$ ). In particular, the component $\eta_3(u=s)$ for a "scalar" electron is given by Eq.(\ref{eta3}) if we omit the item $s^2/(1+s)/2$ in the denominator.

When both terms in (\ref{prosp}) contribute to the radiation, a corresponding alignment is sometimes called string~-~of~-~string (SOS) orientation, since at such an alignment particles traverse axes (strings) forming the plane. In this case, the emission of hard photons is described by the second term in (\ref{prosp}) being CB by nature. A difference of
\begin{figure}[h]
\centering
\includegraphics[width=0.48\textwidth]{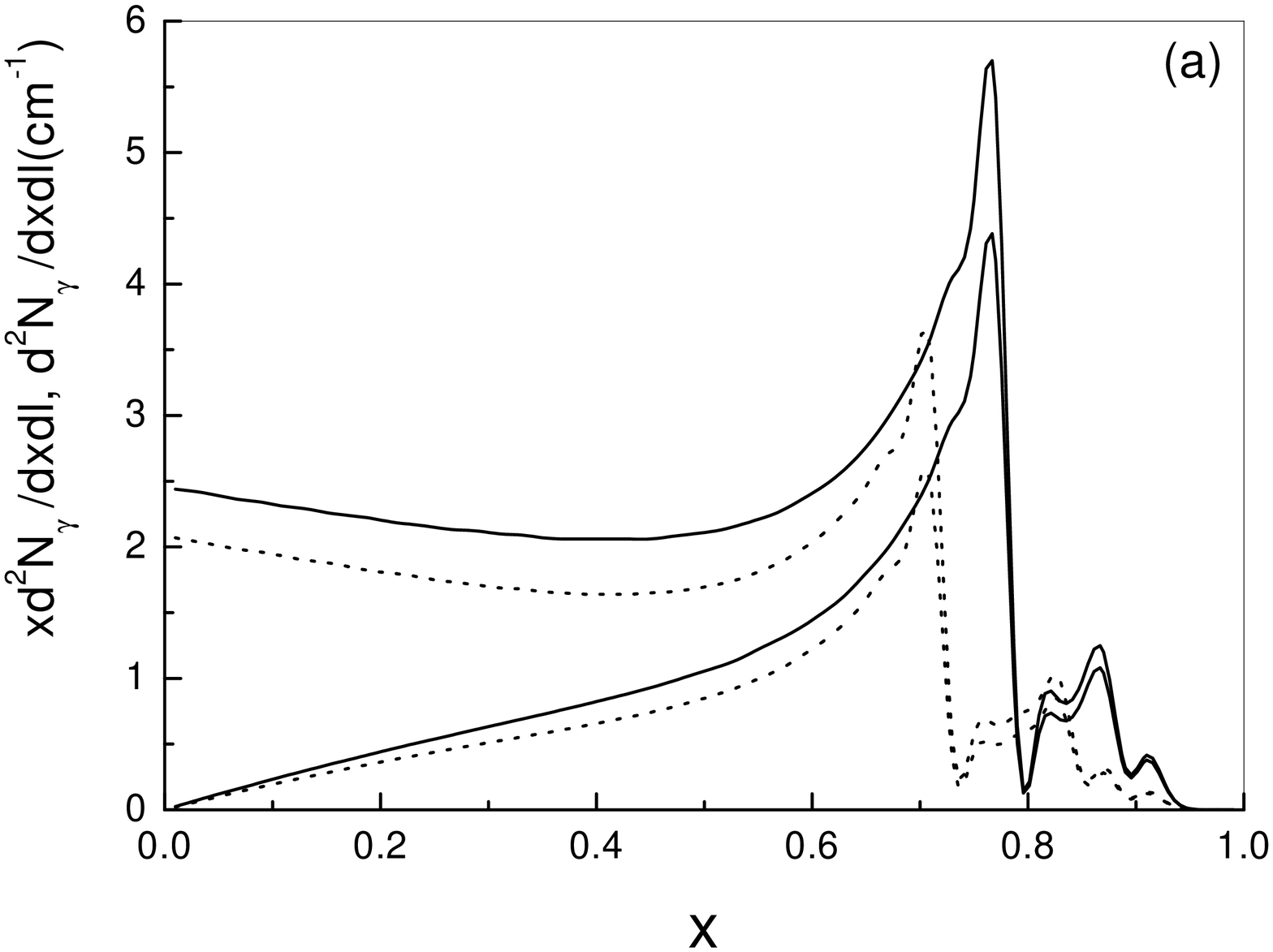}\hspace{0.025\textwidth}
\includegraphics[width=0.48\textwidth]{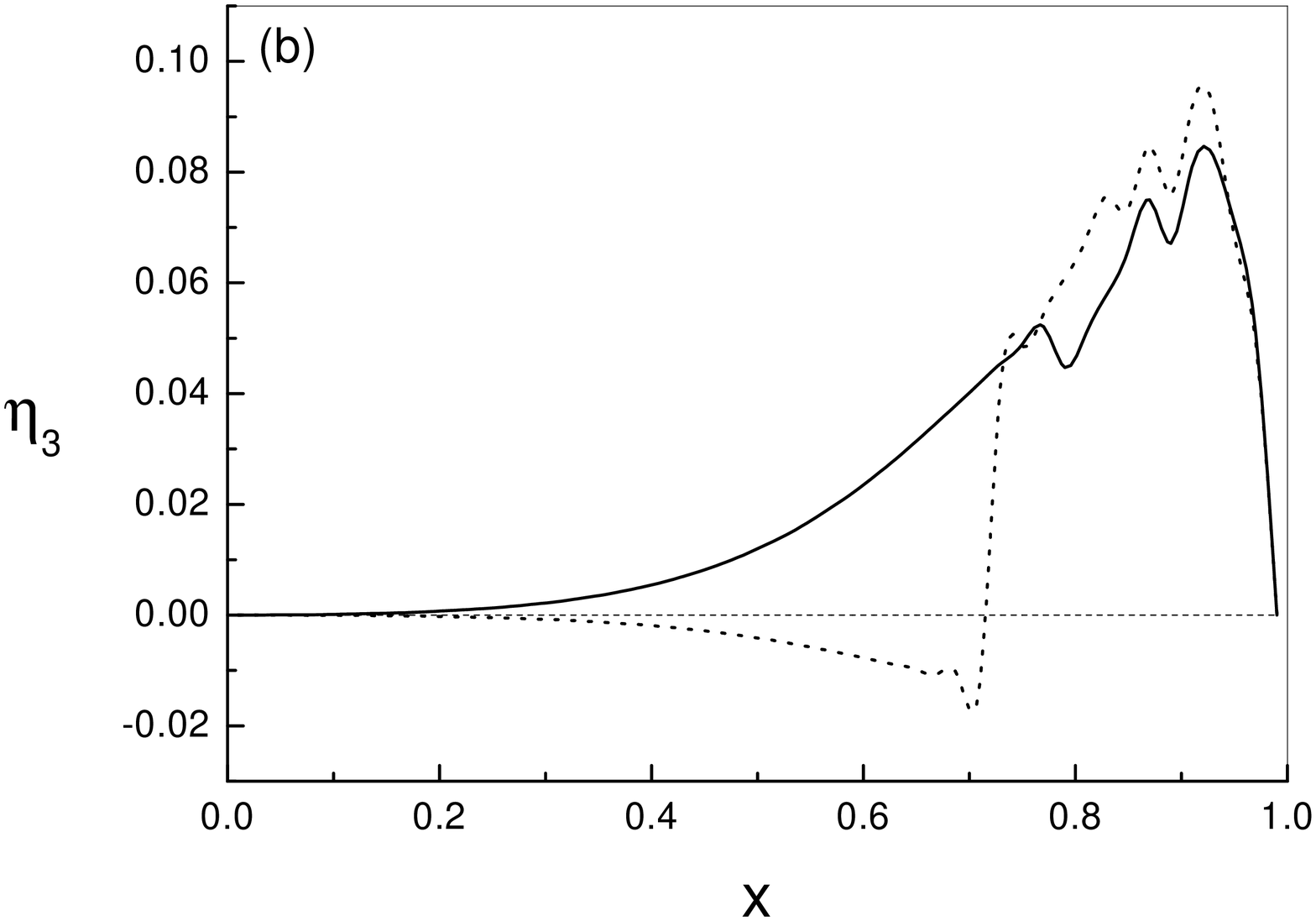}
\caption{(a): Contribution of the CB-like term ( Eq.(\ref{prosp}) at SOS orientation ( 0.3 mrad off the  $<001>$-axis in the $(1\bar{1}0)$-plane  ). Intensities and probabilities in diamond at $\varepsilon=150$ GeV (solid ), and in silicon at $\varepsilon=180$  GeV (dotted);(b): third component of the Stokes vector at the same settings.}\label{Fig:sosins3}
\end{figure}
spectra and polarizations for photons emitted by means of the same physical mechanism (CB) is completely due to that of equivalent photon fluxes at different orientations. The quantity $|q_{\parallel}|$ which determines the shape of a spectrum is independent of $q_2$ at SOS: $q_{\parallel}=q_1\vartheta_0$. Then the summation over $q_1$ in $\sum\limits_{\bm{q}_{\perp}}= \sum\limits_{q_1} \sum\limits_{q_2}$ (see (\ref{ccom})) corresponds to the splitting of the total flux into subsets of equivalent photons having the same $q_{\parallel}$. Remember, that for $x\leq x_{max}$ ($ x_{max}$ marks the first peak position ) the radiation is described within a high accuracy by only one subset having the minimal value of $|q_{\parallel}|$. This value is provided by $|q_1|_{min} = 2\pi/d_{ax}$ where $d_{ax}$ is the distance between axes forming the plane. As the quantity $b^{(C)}_3$ in (\ref{ccom}) is proportional to $q_1^2-q_2^2$, only first ($q_2=0$) term in the sum over $q_2$ is positive for the main ($|q_1|=|q_1|_{min}$) subset. When $q_2$ increases, the magnitude of negative terms diminishes. Their sum, however, cancels almost perfectly the first term. This results in a small magnitude of $\eta_3$ seen in Fig.\ref{Fig:sosins3}. 
For the next ($|q_1|=2|q_1|_{min}$) subset, already two first terms are positive which leads to a positive total sum over $q_2$. As a result, the polarization is somewhat higher for $x>x_{max}$ being parallel to the plane ($\eta_3>0$). Qualitatively, the equivalent photon beam at SOS orientation turns out to be almost unpolarized in contrast to the above example of pure CB. The component $\eta_1$ vanishes at SOS orientation. This can be easily verified if we, e.g., change the sign of $q_2$ in the sum $\sum\limits_{q_2}$ in (\ref{ccom}).

The total ( integrated over $\omega$ ) probabilities, $W_{\gamma}^{tot}$, are typically the order of magnitude larger at SOS orientation mainly due to the PFC. More precisely, we have in the above examples for silicon $W_{\gamma}^{tot}(CB)\simeq 1.2 cm^{-1}$, while at SOS orientation the PFC and CB-like terms give correspondingly $42.9 cm^{-1}$ and $2.2 cm^{-1}$.

To describe a shower development, the probability of $e^+e^-$-pair production by a photon, $dW_e$, is needed as well. Using Eqs.(3.12),(3.25) in \cite{book}, we obtain, first, the expression analogous to (\ref{photen}) where the summation over positron final states and electron spin states has been performed
\begin{equation}\label{paren}
dw_e=\frac{\alpha}{(2\pi)^2}\frac{d^3p}{\omega }\frac {\varepsilon} {2\varepsilon'}\int dt_1 dt_2
L_e(t_1,t_2)\exp[-i\frac {\varepsilon} {\varepsilon'}(k,x_1-x_2)]\,\,.
\end{equation}
Here $\bm{p}$ and $\varepsilon$ are the electron momentum and energy, $\varepsilon'= \omega -\varepsilon $, and
\begin{equation}\label{Lpar}
L_e(t_1,t_2)=(\bm{e}^*\bm{v}_1)(\bm{e}\bm{v}_2)(\varphi( \varepsilon) -2 )+[(\bm{e}^*\bm{e})(\bm{v}_1\bm{v}_2-1+\gamma^{-2})-(\bm{e}^*\bm{v}_2)(\bm{e}\bm{v}_1)](\varphi( \varepsilon) +2 ).
\end{equation}
Note that the quantities $L_e(t_1,t_2)$ in (\ref{Lpar}) and $L(t_1,t_2)$ in (\ref{Ltt}) turn into each other if we change $(\varphi( \varepsilon) -2)\longleftrightarrow(\varphi( \varepsilon) +2 )$. Further consideration of the pair production may be performed using the same approach and approximations as those applied above to the photon emission problem. Here we give explicitly only a perturbation ( CB-like ) term in the expression for $dW_e$ in the form of (\ref{prosp})
\begin{equation}\label{propar}
\frac{dW_e}{d\varepsilon}=\frac{dW^{(F)}_e}{d\varepsilon}+ \frac{dW^{(C)}_e}{d\varepsilon}\,,\quad \frac{dW^{(C,F)}_e}{d\varepsilon} = A^{(C,F)}_e+\bm{B}^{(C,F)}_e\bm{\eta}\,,
\end{equation}  
where $\bm{\eta}$ describes the photon polarization, and
\begin{eqnarray}\label{parcom}
\Bigl ( A^{(C)}_e,\bm{B}^{(C)}_e\Bigr)\!\!&=&\!\!\frac{\alpha}{\omega^2}\sum\limits_{\bm{q}_
{\perp}}^{}\Biggl | \frac{G(\bm{q}_{\perp})\bm{q}_{\perp}}{q_{\parallel}}\Biggr |^2\Bigl ( a^{(C)}_e,\bm{b}^{(C)}_e \Bigr)\theta(1-\beta)\,;\quad \beta= \frac{\omega m^2}{2\varepsilon \varepsilon' |q_{\parallel}|}\,,\nonumber \\\\a^{(C)}_e\!\!&=& \!\!\frac{1}{4}\varphi(\varepsilon)+\beta (1-\beta)\,,b^{(C)}_{e1}=-\beta^2\nu_1\nu_2 \,,b^{(C)}_{e2}=0\,,b^{(C)}_{e3}=-\frac{1}{2}\beta^2(\nu_1^2-\nu_2^2)\,.\nonumber 
\end{eqnarray} 
Note that this result may be obtained from (\ref{ccom}) by means of the substitution mentioned above and evident change in the common multiplier. The plane field contribution to the pair production probability ( first term in (\ref{propar})) was investigated in \cite{Pappla} where the CFA was used. Though the applicability of this approximation is questionable ( see discussion above ), we used the results of \cite{Pappla} as a rough estimate and found that PFC to the pair production probability should be neglected under our conditions.
\section{Radiation from thick crystals}
As long as the crystal thickness, $L$, satisfies the condition $N_{\gamma}\sim W_{\gamma}^ {tot}L \ll 1 $ (thin crystal), the radiation emitted is described by formulas obtained in the previous Section. For such thicknesses, the relative energy loss, $\Delta \varepsilon / \varepsilon_0$, is even smaller than the number, $N_{\gamma}$, of photons emitted. Since the total probability, $W_{\gamma}^{tot}$, depends on the initial electron energy $ \varepsilon_0$ and crystal orientation, the same sample may prove to be thin or thick ($N_{\gamma} \sim 1 $) depending on settings. At the noticeable ($N_{\gamma} \gtrsim 1 $) yield, an alteration of the particle energy can  no longer be neglected, several photons are emitted and the electron-photon shower develops.

The main processes taken into account in our simulation of the $e^+e^- \gamma$-shower development are: i) emission of photons due to the coherent  and incoherent mechanisms, ii) absorbtion of photons due to the $e^+e^- $-pair production by both mechanisms, iii) multiple scattering of electrons and positrons. Many-dimensional maps of probabilities were created describing the photon emission and pair production depending on the energy, momentum direction, and polarization. In other words, thousands distributions like those shown in  Figs.\ref{Fig:plane1},\ref {Fig:cbins2},\ref{Fig:sosins3} have been obtained while a  calculation of the PFC was the most arduous task. Different mechanisms were simulated as independent ones. In particular, each photon emitted by the coherent mechanism was provided with the polarization according to Eq.(\ref{prosp}) and was unpolarized when emitted incoherently. The angular divergence of the initial electron beam was also taken into account. The values for this divergence of 30 $\mu$rad and 50 $\mu$rad used in our calculations correspond to the experimental conditions of \cite{NA59} and \cite{kirsebom} as do also the initial energies  and angles of incidence. Note that the same settings were used in above examples (see Figs.\ref {Fig:cbins2}, \ref{Fig:sosins3}) illustrating instantaneous characteristics of a radiation. So, we can compare the outputs from thin and thick crystals.

One must distinguish the true power spectrum of a radiation from that of energy losses. The latter is observed when a detector (e.g., a calorimeter) sums up over the energies of all photons emitted by one electron. These spectra coincide in the limit of vanishing crystal 
\begin{figure}[h]
\centering
\includegraphics[width=0.48\textwidth]{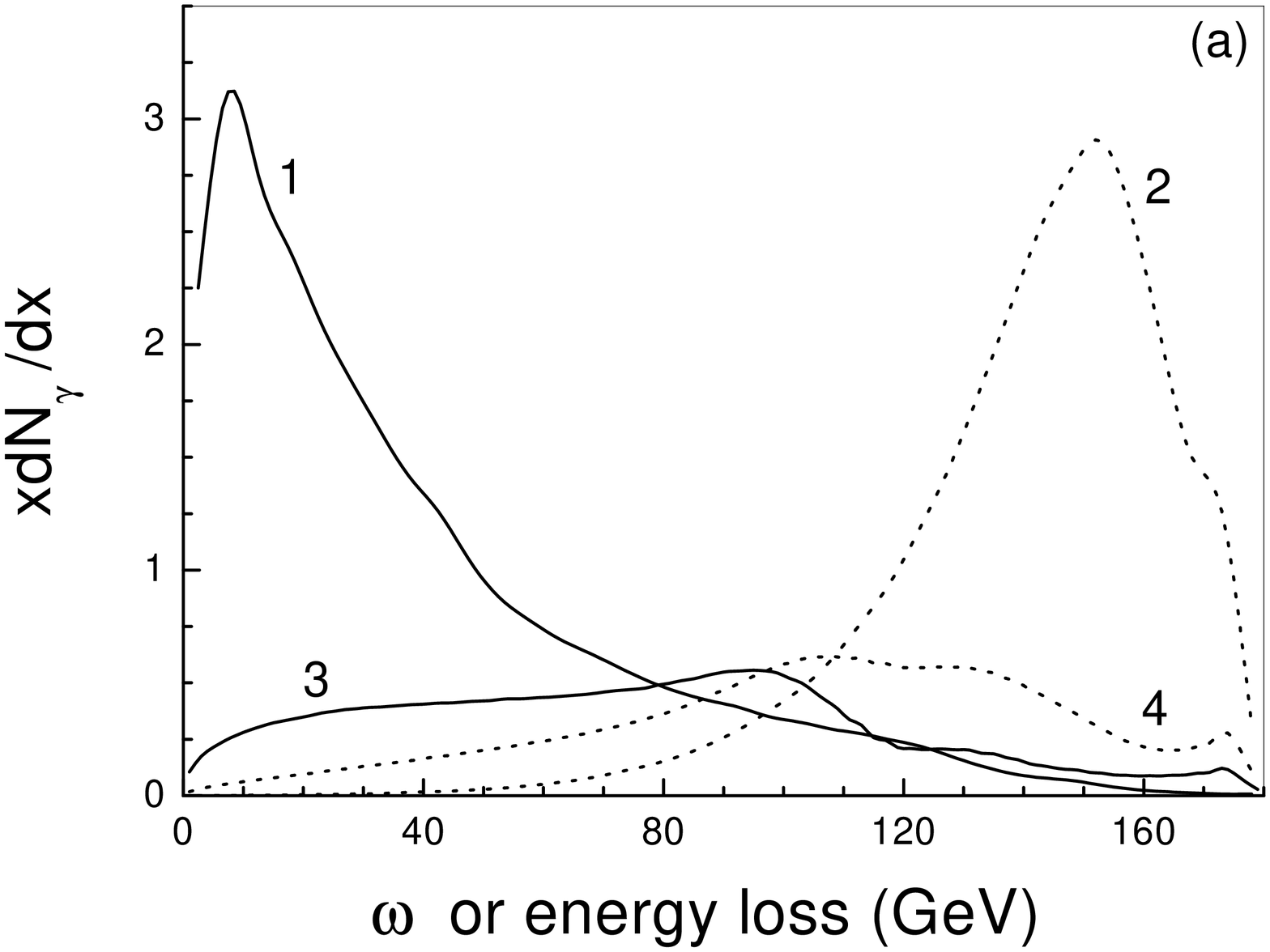}\hspace{0.025\textwidth}
\includegraphics[width=0.48\textwidth]{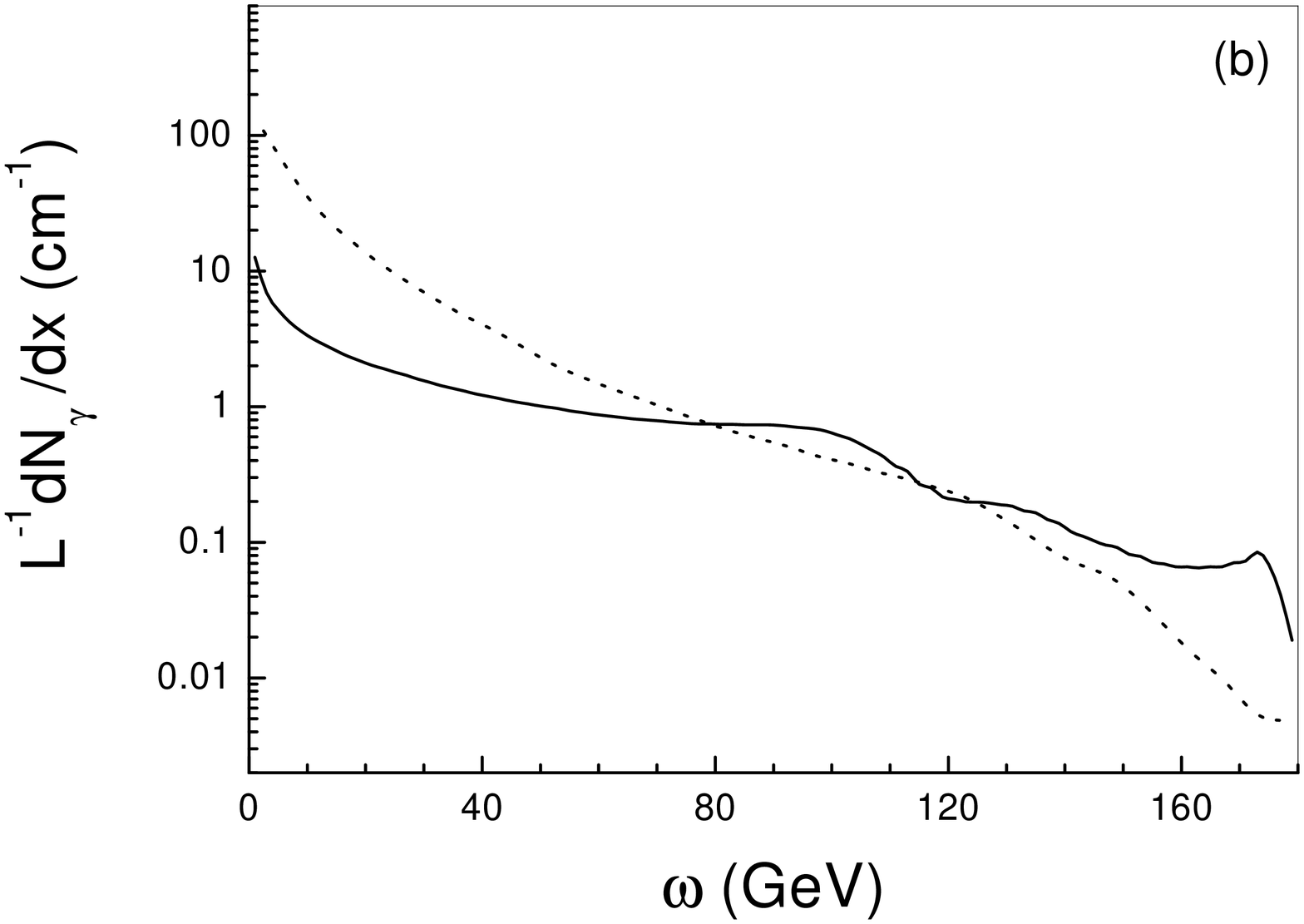}
\caption{Yield from 1.5 cm-thick silicon crystal at (center of beam) settings as in Figs.\ref{Fig:cbins2},\ref{Fig:sosins3};(a): Power and energy loss spectra for SOS (1,2) and CB (3,4); (b): spectra (effective probabilities) for SOS (dotted) and CB (solid).}\label{Fig:shinmul4}
\end{figure}
thickness when the multiplicity (a number of photons at given nonzero total energy loss) tends to unity. In our case, they are very different, especially for SOS orientation (cf. curves 1 and 2 in Fig.\ref {Fig:shinmul4}a) when the PFC is dominant leading to rather soft power spectrum. Note that due to the same mechanism (PFC) of photon emission, the multiplicity is very high just for SOS orientation (cf. curves 1 and 3 in Fig.
\ref{Fig:shspol5}b). Remember that the multiplicity increases when the lower energy boundary of photon  recorded , $\omega_{th}$, decreases. Results presented in Fig.\ref{Fig:shspol5}b were obtained at $\omega_{th}=1\,$GeV. In applications, power spectra are less interesting than the distributions in number of photons (spectra). Such spectra are presented in Fig.\ref {Fig:shinmul4}b in the form of effective probabilities capable of direct comparison with probabilities shown in
Figs.\ref {Fig:cbins2}, \ref{Fig:sosins3}. The shower spectra are significantly softer than initial ones due to the decrease of the mean energy of charged particles with the increasing depth, and to the incoherent mechanism action. Recollect, that the parameter $|q_{\parallel} |$ which determines the position of hard peaks in the instantaneous spectrum depends on the current energy and velocity direction. The latter also changes in thick crystals mainly due to the multiple scattering. Note that for conditions of Fig.\ref {Fig:shinmul4}, the mean-square scattering angle ( at the initial energy ) is about 50 $\mu$rad being larger than the angular divergence of the initial electron beam. As a result, a smearing of peaks takes place and sharp structures are not seen in shower spectra. All the factors mentioned
\begin{figure}[h]
\centering
\includegraphics[width=0.48\textwidth]{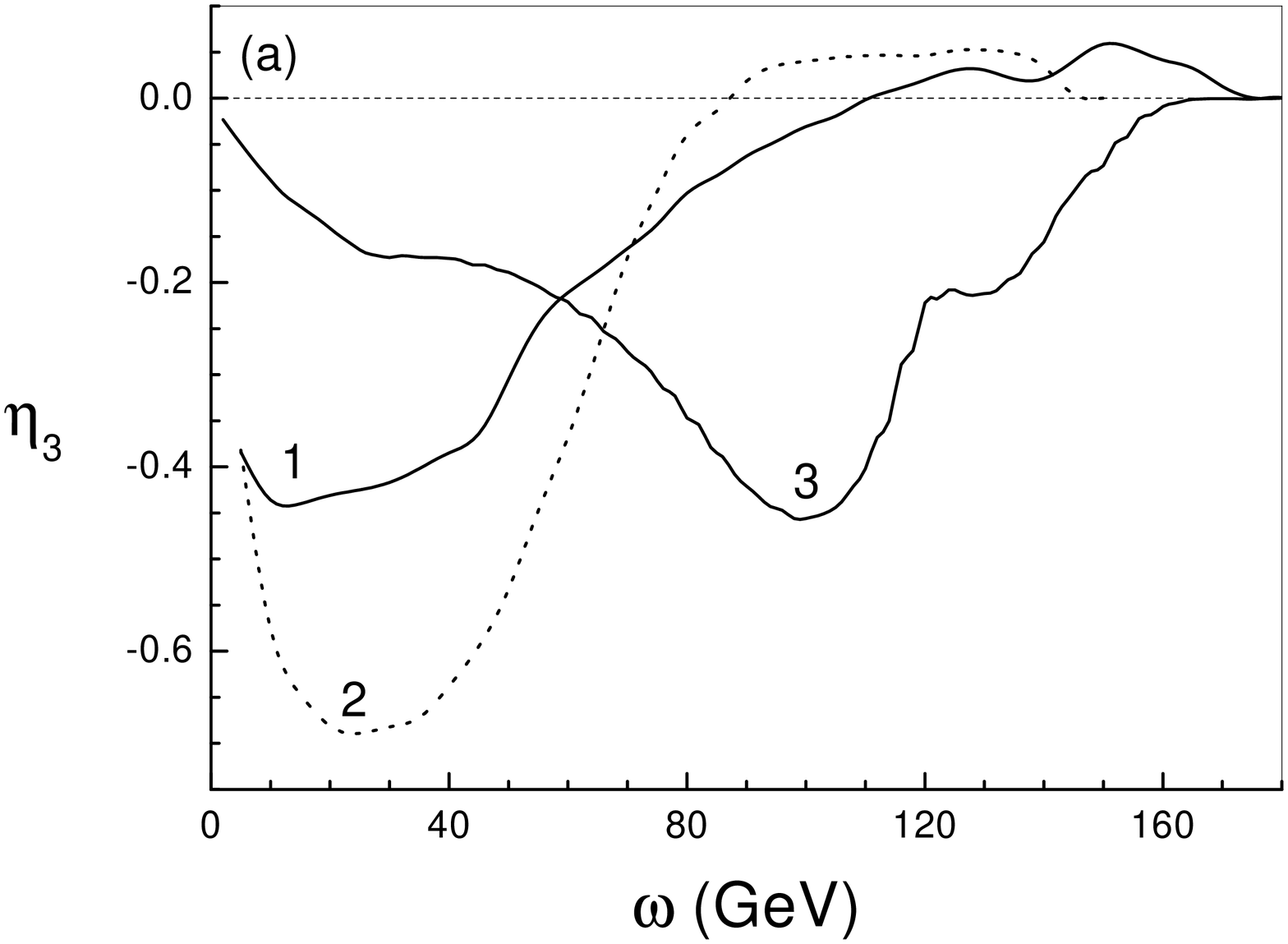}\hspace{0.025\textwidth} 
\includegraphics[width=0.48\textwidth]{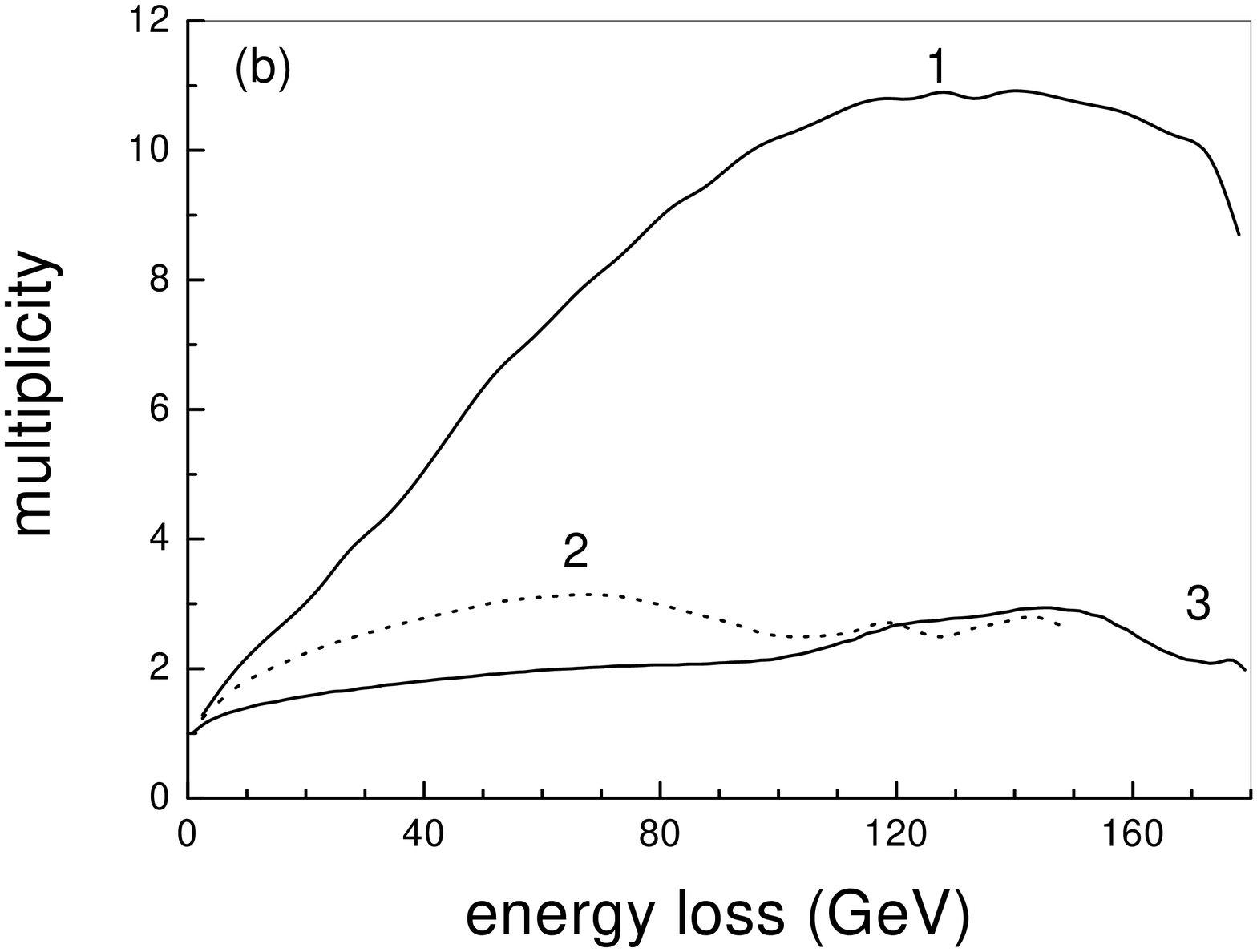}
\caption{(a): Polarization ($\eta_3$) at settings of Fig.\ref{Fig:shinmul4} for SOS (1), CB (3), and for SOS from 0.05 cm-thick diamond crystal (2) at settings of Fig.1 in \cite{kirsebom}( $\varepsilon_0=150$ GeV, $\vartheta_0=$ 0.3 mrad $|\psi|\le 10\mu$rad)
; (b): multiplicities at these conditions for $\omega_{th}=1\,$GeV.}\label{Fig:shspol5}
\end{figure}
affect the shape of polarization distributions shown in Fig.\ref{Fig:shspol5}a. For CB ( see curve 3 in Fig.\ref{Fig:shspol5}a ), this distribution is maximal at $\omega\simeq 100$ GeV. The shift left (by $\sim 11$ GeV as compared to Fig.\ref{Fig:cbins2}b) is due to the energy and angular spread mentioned above, while a diminution of the magnitude is mainly caused by the incoherent (unpolarized photons) contribution. For evident reasons, such changes are marked feebly for relatively thin diamond ( see curve 2 in Fig.\ref{Fig:shspol5}a ) crystal.

\begin{figure}[h]
\centering
\includegraphics[width=0.48\textwidth]{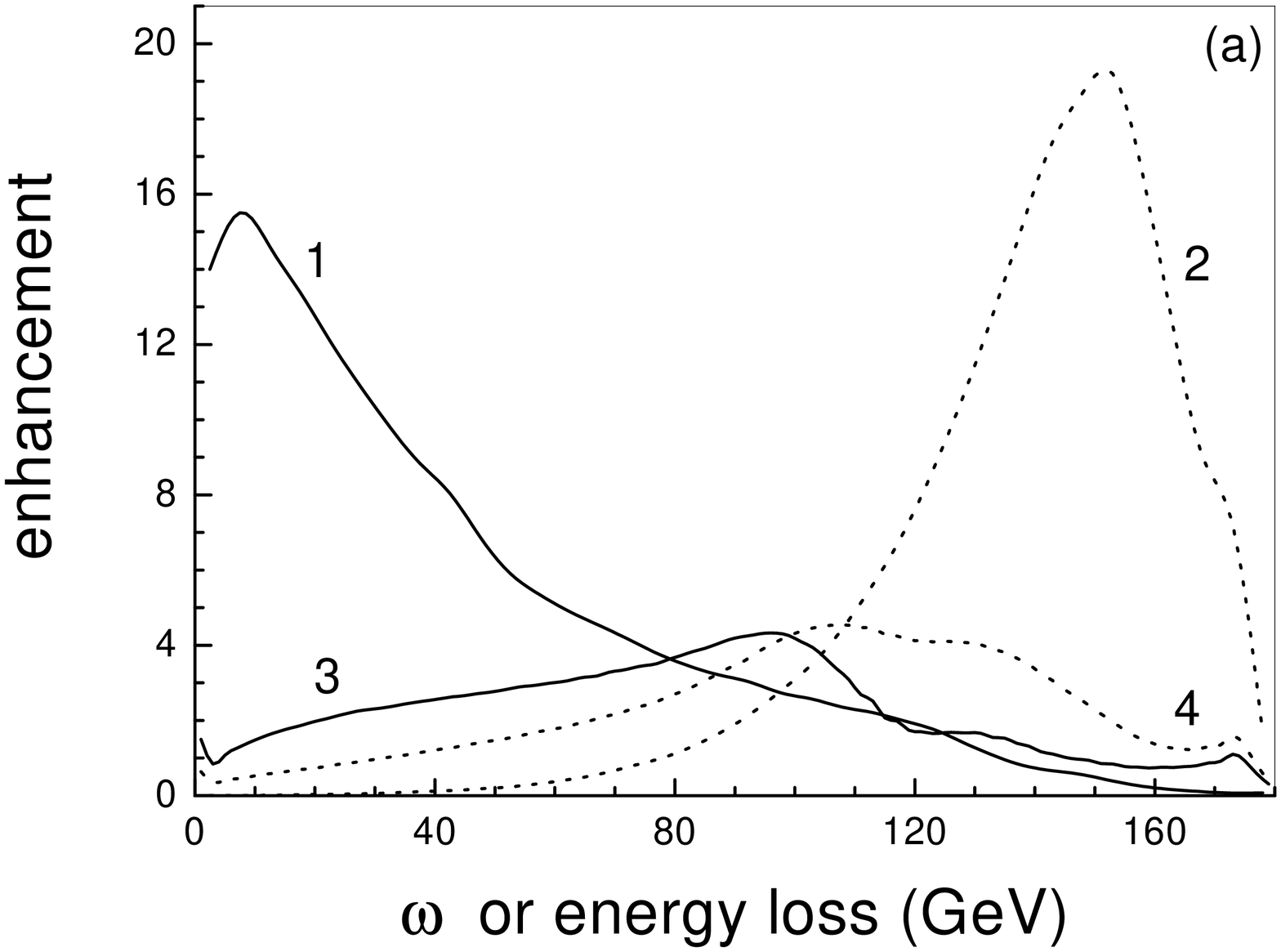}\hspace{0.025\textwidth}
\includegraphics[width=0.48\textwidth]{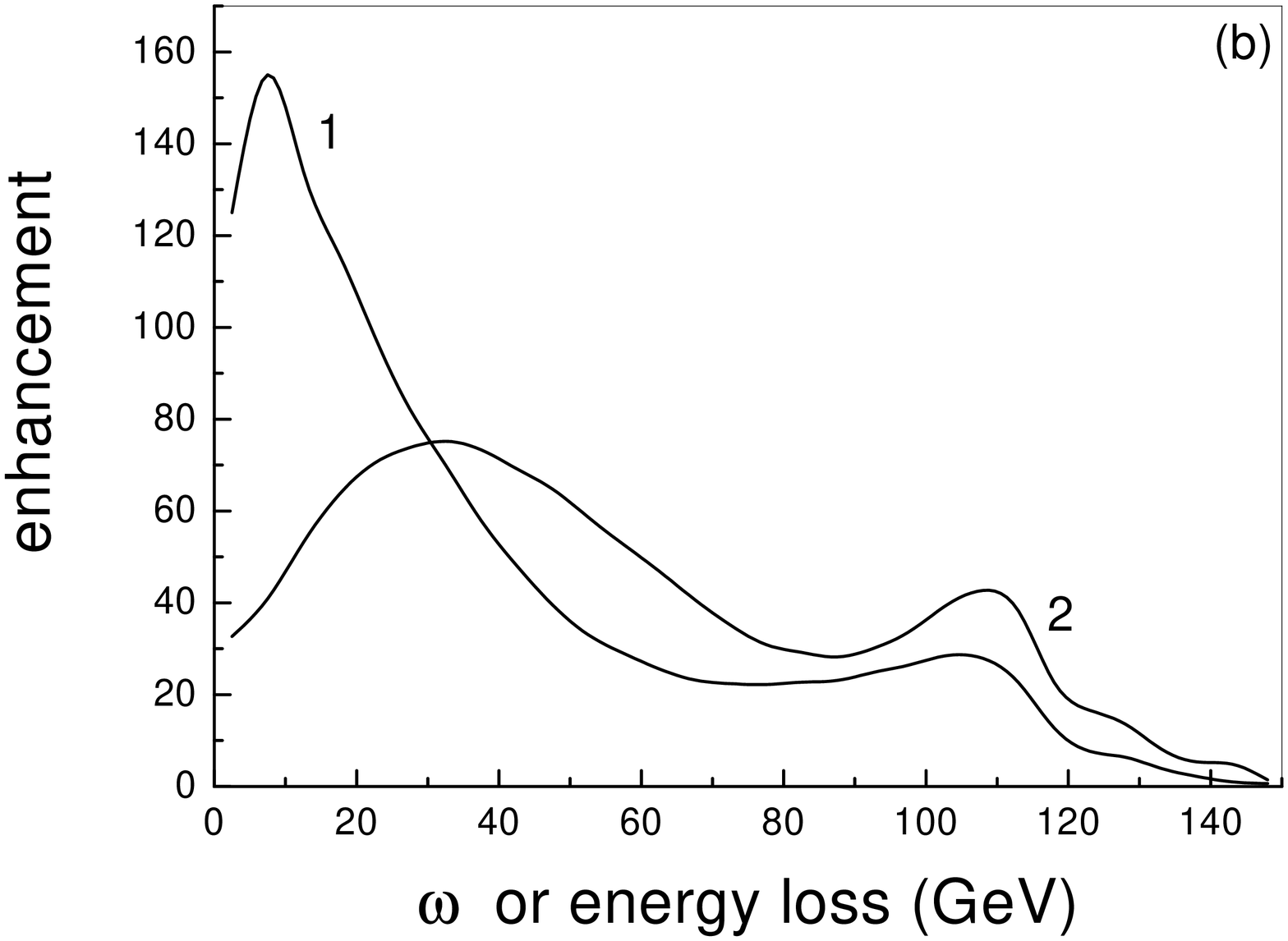}
\caption{ At settings of Fig.\ref{Fig:shspol5}; (a): probability and energy loss enhancements in silicon for SOS (1,2) and CB (3,4); (b): the same for SOS in diamond}\label{Fig:enhan6}
\end{figure}

Let us define an enhancement as bin-by-bin ratio of the yields from oriented to disoriented crystals. In the latter case, only incoherent (amorphous-like) mechanism is acting, when $dN_{\gamma}^{(am)}/dx \sim Q(x)/x$ with $Q(x)$ being rather smooth function of $x$. Therefore, the ratio of numbers of photons ( probability enhancement ) is very like in the form to a power spectrum while the energy loss enhancement bears a strong resemblance to the energy loss spectrum shape (cf. corresponding curves in Figs.\ref{Fig:shinmul4}a and \ref{Fig:enhan6}a ). Note that an enhancement increases with decreasing thickness $L$, other things being equal. This explains, along with a larger bare (at $L\rightarrow 0$) probability enhancement for diamond, the order of magnitude difference in these quantities for silicon and diamond crystals (cf. curves 1 in Figs. \ref{Fig:enhan6}a and \ref{Fig:enhan6}b ).
Already from Figs. \ref{Fig:shinmul4},\ref{Fig:shspol5}, the SOS orientation looks less favorable than CB for the hard photon production. As explained above, this is due to the PFC which in itself is characterized by relatively soft spectra with large intensities and total probabilities. So that, the CB-like contribution providing a hard photon emission is suppressed at SOS, in particular, due to the energy loss via competing mechanisms. However, 
\begin{figure}[h]
\centering
\includegraphics[width=0.6\textwidth]{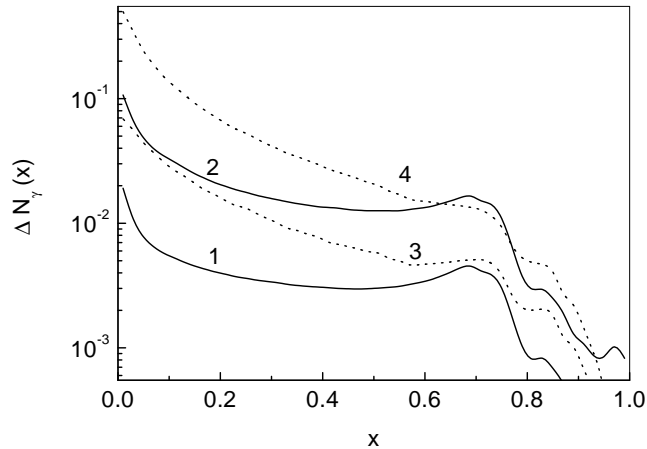}
\caption{ Number of photons per bin $\Delta x=0.02\,$ in a diamond crystal at $\varepsilon_0=250 GeV $ for CB ( 5 mrad off the $<001>$-axis and 160 $\mu$rad off the $(1\bar{1}0)$-plane ) at $L=0.1 $cm (1) and $L= 0.5$ cm (2); for SOS  (  325 $\mu$rad off the $<110>$-axis in the (001)-plane ) at $L=0.04$ cm (3) and $L=0.2$ cm (4); $x=\omega/\varepsilon_0$.}\label{Fig:gamco7}
\end{figure}
the thickness in above examples was chosen to optimize the yield of CB at $\omega \sim 100\,$GeV being not optimal for SOS. Additionally, the positions of peaks (cf. Figs. \ref {Fig:cbins2} and \ref{Fig:sosins3} ) in the initial spectra were different for two orientations. Let us now compare the yield from a diamond crystal at $\varepsilon_0= 250\,$GeV for three different orientations characterized by the same peak position in instantaneous spectra, $x_{max}=0.75$. Those are two SOS orientations ( 325 $\mu$rad off the $<110>$-axis in the (001)-plane and  180 $\mu$rad off the $<001>$-axis in the $(1\bar{1}0)$-plane ) and CB ( 5 mrad off the $<001>$-axis and 160 $\mu$rad off the $(1\bar{1}0)$-plane ). From Fig. \ref{Fig:plane1}, the weaker (001)-plane is preferable to the stronger  $(1\bar{1}0) $-plane for the purpose of hard photon production. Additionally, stronger $<110>$-axes are involved in the first of the two SOS orientations increasing the CB-like contribution. As a result, the hard ($x > 0.5\,$) photon yield turned out to be higher for this orientation at any crystal thickness. For example, this is 1.5 times as large in the region $x \sim 0.6\div 0.75$ at $L=0.2$ cm. Just this orientation is compared to pure CB in Fig. \ref{Fig:gamco7}. For CB, a local maximum in the shower spectra situated at $x = 0.69\,$ is seen even at larger $L$ than those presented in Fig. \ref{Fig:gamco7}. At SOS, such a maximum (at $x = 0.71\,$) is feebly marked already at $L=0.04 $ cm, so that the spectra are monotonically decreasing for larger $L$. However, an increase of the yield at $x = 0.71\,$ continues up to $L=0.3$ cm where it amounts only to $3\%$ of the yield at $L=0.2$ cm. Thus, a saturation occurs in the hard part of the SOS-spectrum at thicknesses $L\sim 0.2$ cm. In our examples of CB, the thicknesses are chosen so as being noticeably smaller than those leading to the saturation of CB-spectra, they provide almost the same amount of hard photons as at SOS. Even under these conditions ( the hard photon yield of CB may be further increased ), CB-spectra would get the better in applications being on the whole much harder than SOS-spectra (cf. curves 1,3 and 2,4 in Fig.\ref {Fig:gamco7}), and having a polarization ( about 40$\%$ near $x_{max}$). Numerous relatively soft photons may provide rather severe  background conditions, producing, e.g., $e^+e^-$-pairs or hadrons directly in a radiator.

In conclusion, we hope that explicit formulas presented along with a qualitative analysis performed allow anyone to make own estimations of radiation characteristics at any orientation where polarized photons may be obtained from unpolarized electrons or positrons penetrating through single crystals.
\appendix
\renewcommand{\thesection}{\appendixname\hspace{0.5em}\Alph{section}}
\renewcommand{\theequation}{\Alph{section}.\arabic{equation}}
\renewcommand{\thetable}{\Alph{section}.\arabic{table}}
\setcounter{equation}{0}
\section{}
\label{AppendixA}
Let x be the coordinate perpendicular to some system of crystal planes with the inter-planar distance, $d_{pl}$. Then the periodic plane potential for electrons reads
$$ U_e(x)=-\sum\limits_{n=-\infty}^{\infty}G(nq)\exp(i\pi ny)\,,\quad q=\frac{2\pi}{d_{pl}} \,,\quad y=\frac{2x}{d_{pl}}\,.$$ As the direct use of this potential is impossible  in analytical calculations, several approximate forms ( see Chapter 9,15 in \cite{book} ) were suggested. Being not satisfied with previously used forms, we propose here the new one which provides a precise fit for any crystal plane potential and very simple expressions for the
velocity Fourier-transforms. For $0 \le y\le 1$, it reads  
\begin{equation}\label{potel}
U_e(y)=-U_{pl}[\theta
(y_1-y)(1-a_1y^2)+\theta (y-y_1)\theta (y_2-y)(a_2y^2+by+c)+\theta
(y-y_2)a_3(1-y)^2]\,\,,
\end{equation}
where $\theta(x)$ was defined in (\ref{ccom}) being the step function, $U_{pl}$ is the potential well depth. The origin is set to the point where the potential is minimal ($U_e(0) =-U_{pl}$),i.e., just at the plane. Beyond the segment $y\in [0,1]$, the values of $U_e(y)$ may be obtained from (\ref{potel}) using evident symmetry and periodicity conditions. For positrons, we also choose the origin at the point where the corresponding potential, $U_p(y)$, is minimal ($U_p(0) =0$),i.e., in the middle between two neighboring planes. So, for $0 \le y\le 1$, we have $U_p(y)=-U_e(1-y)$ with $U_e(y)$ defined in (\ref{potel}). Note
that only three of seven fitting parameters $ a_i$, $y_i $,$b$, and $c$ in (\ref{potel}) are independent. Using continuity conditions of the potential and corresponding electric field in points $y_1$,$y_2$, we can, for example, express the rest parameters via $ y_1$, $a_1 $, and $a_2 $:
\begin{eqnarray}\label{param}
a_3&\!=&\!\frac{a_2-y_1^2a_1(a_1+a_2)}{1-a_1+(1-y_1)^2 (a_1+a_2)}\,,\quad y_2=\frac{y_1(1-y_1)(a_1+a_2)-1}{a_2-y_1(a_1+a_2)}\,,\nonumber\\ \\ b&\!=&\!-2y_1(a_1+a_2)\,,\quad c=1+y_1^2(a_1+a_2) \,\,.\nonumber
\end{eqnarray}
Independent fitting parameters along with $ U_{pl}$ and $ d_{pl}$ are listed in Table \ref{partab} for some crystal planes.
\begin{table}[h]
\centering
\caption{Potential well depths, $U_{pl}(eV)$, inter-planar distances, $d_{pl}(\AA)$, and parameters of the fit (\ref{potel}) for some crystal planes.}
\begin{tabular}{|c|c|c|c|c|c|}
\hline
\multicolumn{1}{|c|}{Crystal, plane}&$ U_{pl}$&$ d_{pl}  $&$ y_1$&$a_1 $&$a_2 $\\
\hline
 diamond (110)& 23.54&1.261&0.109&10.57&1.994\\ \hline
 diamond (001)&12.06&0.892&0.152&7.16&1.623\\ \hline 
 silicon (110)&21.27&1.920&0.100&13.45&2.750\\ \hline
 silicon (001)&11.73&1.358&0.140&8.63&2.018\\ \hline
 copper (110)&34.14&1.278&0.161&7.00&1.660\\ \hline
 iron (110)&68.88&2.027&0.084&17.38&3.104\\ \hline
 tungsten (110)&132.69&2.238&0.054&34.90&5.656\\ \hline 
\end{tabular}
\label{partab}
\end{table}

The equation of one~-~dimensional motion in the potential $U(y)$ is
\begin{equation}\label{motion}
\dot{y}=\pm\kappa\sqrt{z-U(y)/U_{pl}}\,,\quad \kappa=2\theta_{pl}/d_{pl}\,,\quad \theta_{pl} =\sqrt{2U_{pl}/\varepsilon}\,\,,
\end{equation}
where  $z=\varepsilon_{\perp}/U_{pl}=U(y)/U_{pl}+(\dot{y}/\kappa)^2$ is the integral of motion (transverse energy in units of $U_{pl}$). Remember now that the integration over time in formulas describing a radiation ( see Section 1) is performed at fixed value of $z$ when, by means of a time shift, any initial conditions may be reduced to the standard one $x(0)=0$ providing the condition $v_x(-t)=v_x(t)$. Then we have $v_n=v_{-n}=v_n^*$ in the velocity fourier series $$v_x(t)=\sum\limits_ {n=-\infty}^{\infty}\,v_n \exp(in\omega_0 t)\,,$$ where  $\omega_0= 2\pi/T$ is the frequency of motion and $T$ being its period. Finally,
\begin{equation}\label{furgen}
v_n=\frac{2}{T} \int \limits_{0}^{T/2}dt v_x(t)\cos(n\omega_0 t)\,\,.
\end{equation}
There is an additional symmetry, $v_x(t \pm T/2)=-v_x(t)$, for channelled particles when Eq.(\ref{furgen}) passes into
\begin{equation}\label{furche}
v_n^{ch}=\sin\Bigl(\frac{n \pi}{2}\Bigr)\frac{4}{T} \int \limits_{0}^{T/4}dt v_x(t)\sin(n\omega_0 t)\,\,.
\end{equation} 
From this equation, even harmonics vanish at channelling (at $-1\le z\le 0$ for electrons and at $0\le z\le 1$ for positrons). Let us define a quantity, $\tilde{v}_n$,  which is related to $v_n$ by $v_n=\tilde{v}_n \theta_{pl}/g(z)$ where $g(z)= \kappa T/4$ at channelling and $ g(z)= \kappa T/2 $ at over-barrier motion. Solving Eq.(\ref{motion}) and taking elementary integrals in  Eqs.(\ref{furgen}),(\ref{furche}), we obtain for electrons
\begin{eqnarray}\label{vmelec1}
\tilde{v}_n^{(e)}&\!=&\!\theta(z_{1e}-z)\frac{\pi}{4}\sqrt{\frac{z+1}{a_1}}\Bigl(\delta_{n,1}
+\delta_{n,-1} \Bigr)+\frac{1}{Q_e^2+a_2}\Biggl \{\theta(z-z_{1e})\frac{a_1+a_2}{Q_e^2-a_1}\Bigl[\sqrt{z-z_{1e}}Q_e\sin\Psi_{1e}\nonumber\\\\ &\!-&\!a_1 y_1\cos\Psi_{1e}\Bigr]+ \theta(z-z_{2e})\frac{a_3-a_2}{Q_e^2+a_3}\Bigl
[\sqrt{z-z_{2e}}Q_e\sin\Psi_{2e}- a_3(1- y_2)\cos\Psi_{2e}\Bigr] \Biggr\}\,, \nonumber
\end{eqnarray}
where $z_{1e}=U_e(y_1)/U_{pl}=a_1 y_1^2-1\,$, $z_{2e}=U_e(y_2)/U_{pl}=-a_3(1- y_2) ^2\,$, $-1<z_{1e}<z_{2e}<0\,$, and 
\begin{eqnarray}\label{vmelec2}
g_e(z)&\!=&\!\theta(z_{1e}-z)\frac{\pi}{2\sqrt{a_1}}+\theta(z-z_{1e})\Biggl[ \frac{1}{\sqrt{a_1}}\arcsin\Biggl(\sqrt{\frac{1+z_{1e}}{1+z}}\Biggr)+\ln \Biggl(\frac{a_1  y_1+\sqrt{a_2(z-z_{1e})}}{\sqrt{|a_2(z+c)-b^2/4|}} \Biggr)\nonumber\\\nonumber\\
&\!\times&\! \frac{\theta(z_{2e}-z)}{\sqrt{a_2}} \Biggr]+\theta(z-z_{2e})\Biggl[ \frac{1}{\sqrt{a_2}}\ln\Biggl(\frac{a_1 y_1+\sqrt{a_2(z-z_{1e})}}{a_3(1-y_2)+\sqrt{a_2(z-z_{2e})}} \Biggr)+ \nonumber \\\\ &\!+&\!\frac{1}{\sqrt{a_3}} \ln\Biggl(\frac{(1-y_2)\sqrt{a_3}+\sqrt{z-z_{2e}}}{\sqrt{|z|}} \Biggr)\Biggr]\,,\quad\Psi_{1e}=\frac{Q_e}{\sqrt{a_1}}\arcsin\Biggl(\sqrt{\frac{1+z_{1e}}{1+z}}
\Biggr)\,, \nonumber \\\nonumber\\ \Psi_{2e}&\!=&\!Q_e\Biggl[g_e(z)-\frac{1}{\sqrt{a_3}} \ln\Biggl(\frac{(1-y_2)\sqrt{a_3}+\sqrt{z-z_{2e}}}{\sqrt{|z|}} \Biggr)\Biggr]\,,\quad
Q_e=\frac{\pi n}{2g_e(z)}[1+\theta(z)]\equiv \frac{n\omega_0^{(e)}}{\kappa}\,.\nonumber         
\end{eqnarray}
Analogously, we find for positrons
\begin{eqnarray}\label{vmpos1}
\tilde{v}_n^{(p)}&\!=&\!\theta(z_{1p}-z)\frac{\pi}{4}\sqrt{\frac{z}{a_3}}\Bigl(\delta_{n,1}
+\delta_{n,-1} \Bigr)+\frac{1}{Q_p^2-a_2}\Biggl \{\theta(z-z_{1p})\frac{a_3-a_2}{Q_p^2-a_3}\Bigl[\sqrt{z-z_{1p}}Q_p\sin\Psi_{1p}\nonumber\\\\ &\!-&\!a_3(1- y_2)\cos\Psi_{1p}\Bigr]+ \theta(z-z_{2p})\frac{a_1+a_2}{Q_p^2+a_1}\Bigl
[\sqrt{z-z_{2p}}Q_p\sin\Psi_{2p}- a_1 y_1\cos\Psi_{2p}\Bigr] \Biggr\}\,, \nonumber
\end{eqnarray}
where $z_{1p}=-z_{2e}\,$, $z_{2p}=-z_{1e}\,$, $0<z_{1p}<z_{2p}<1\,$, and 
\begin{eqnarray}\label{vmpos2}
g_p(z)&\!=&\!\theta(z_{1p}-z)\frac{\pi}{2\sqrt{a_3}}+\theta(z-z_{1p})\Biggl[ \frac{1}{\sqrt{a_3}}\arcsin\Biggl(\sqrt{\frac{z_{1p}}{z}}\Biggr)+\arccos \Biggl(\frac{a_3(1-y_2)}{\sqrt{a_2(z-c)+b^2/4}} \Biggr)\nonumber\\\nonumber\\
&\!\times&\! \frac{\theta(z_{2p}-z)}{\sqrt{a_2}} \Biggr]+\theta(z-z_{2p})\Biggl[ \frac{1}{\sqrt{a_2}}\arctan\Biggl(\frac{a_1 y_1\sqrt{a_2(z-z_{1p})}-a_3(1-y_2)\sqrt{a_2(z-z_{2p})}}{a_1a_3 y_1(1-y_2)+a_2\sqrt{(z-z_{1p})(z-z_{2p})}} \Biggr)+ \nonumber \\\\ &\!+&\!\frac{1}{\sqrt{a_1}} \ln\Biggl(\frac{y_1\sqrt{a_1}+\sqrt{z-z_{2p}}}{\sqrt{|1-z|}} \Biggr)\Biggr]\,,\quad\Psi_{1p}=\frac{Q_p}{\sqrt{a_3}}\arcsin\Biggl(\sqrt{\frac{z_{1p}}{z}}
\Biggr)\,, \nonumber \\\nonumber\\ \Psi_{2p}&\!=&\!Q_p\Biggl[g_p(z)-\frac{1}{\sqrt{a_1}} \ln\Biggl(\frac {y_1\sqrt{a_1}+\sqrt{z-z_{2p}}}{\sqrt{|1-z|}}\Biggr)\Biggr]\,,\quad
Q_p=\frac{\pi n}{2g_p(z)}[1+\theta(z-1)]\equiv \frac{n\omega_0^{(p)}}{\kappa}\,. \nonumber         
\end{eqnarray}
As expected, differences in electron and positron motion diminish when $z$ (transverse energy) increases. We obtain from above formulas for $z\gg 1$ (high above the potential barrier when $z \simeq (\psi/\theta_{pl})^2$ where $\psi$ is the angle of a particle velocity w.r.t. the plane ) $g_e \simeq g_p \simeq 1/\sqrt{z}$ and therefore $\omega_0^{(e)} \simeq \omega_0^{(p)} \simeq \pi \kappa \sqrt{z}\simeq 2\pi\psi/d_{pl}\equiv q \psi$. From Eq.(\ref{vmelec1}) we find for $z\gg 1$ $$\tilde{v}_n^{(e)}\simeq \frac{1}{(\pi n)^3}\bigl[(a_1+a_2)\sin(\pi ny_1)+(a_3-a_2)\sin(\pi ny_2)\bigr]=-\frac{1}{2zU_{pl}}\int \limits_{0}^{1}dy\cos(\pi ny)U_e(y).$$ So, the velocity Fourier transform is expressed via Fourier coefficients of a potential. This happens when the rectilinear trajectory approximation is valid. If we substitute the original (periodic) plane potential for $U_e(y)$ in this integral, $\tilde{v}_n^{(e)}$ takes the form $\tilde{v}_n^{(e)}\simeq G(nq)/(2zU_{pl})$. Then
the quantity $\gamma v_n$ appearing in (\ref{c1small}) reads: $\gamma v_n\simeq G(nq)/ (m \psi)$. At $z\gg 1$ , the velocity Fourier transform for positrons differs from that for electrons by the factor, $-\cos(\pi n)$, which does not lead to some changes in (\ref{c1small}) as only $v_n^2$ enters into this formula.

\end{document}